%% file: THDM_paper_v3.tex
\documentclass[11pt,a4paper]{article}
\usepackage{jheppub}
\usepackage[utf8]{inputenc}
\usepackage{amsmath}
\usepackage{amsfonts}
\usepackage{amssymb}
\usepackage{graphicx}
\usepackage{float}
\usepackage{cancel}
\usepackage{extarrows}
\usepackage{caption}
\usepackage{subfig}
\usepackage{rotating}
\usepackage[toc,page]{appendix}
\usepackage{cleveref}

\graphicspath{{.}{images/}}

\usepackage{caption}
\makeatletter

\newcommand{\Rmnum}[1]{\expandafter\@slowromancap\romannumeral #1@}
\makeatother

\title{High Scale Boundary Conditions in Models with Two Higgs Doublets.}
\author{John McDowall,}
\author{David J. Miller}

\affiliation{SUPA, School of Physics and Astronomy, University of Glasgow, \\
						Glasgow, G12 8QQ, United Kingdom}

\emailAdd{j.mcdowall.1@research.gla.ac.uk}
\emailAdd{david.j.miller@glasgow.ac.uk}

\abstract{
We investigate high scale boundary conditions on the quartic Higgs-couplings and their $\beta$-functions in the Type-II Two Higgs Doublet Model and the Inert Doublet Model. These conditions are associated with two possible UV physics scenarios: the Multiple Point Principle, in which the potential exhibits a second minimum at $M_{Pl}$, and Asymptotic Safety, where the scalar couplings run towards an interacting UV fixed point at high scales. We employ renormalisation group running at two-loops and apply theoretical and experimental constraints to their parameter spaces. We find neither model can simultaneously accommodate the MPP whilst also providing realistic masses for both the Higgs and the top quark. However, we do find regions of parameter space compatible with Asymptotic Safety.}

\begin{document}
\maketitle
\section{Introduction}
\label{sec:THDM_introduction}
The discovery of the Higgs at ATLAS \citep{Aad:2012tfa} and CMS \citep{Chatrchyan:2012xdj} supports the Standard Model's (SM) mechanism of breaking the $SU(2) \times U(1)$ electroweak symmetry, which requires a single $SU(2)$ complex scalar doublet and results in one neutral scalar particle. The simplicity of the SM scalar sector is striking given the complexity of its fermion sector, so it's no surprise that the notion of extending the SM with additional scalar fields has motivated much of modern particle physics research.

In a recent work we looked at the possibility of high scale boundary conditions arising in the complex singlet extension of the SM \citep{McDowall:2018tdg}. This was motivated in part by the very small value of both the SM Higgs quartic coupling $\lambda$ and its $\beta$-function $\beta_{\lambda}$ at the Planck scale $M_{Pl}$. The possibility that this interesting feature of the SM is a high scale boundary condition derived from additional physics at $M_{Pl}$, and its consequences for e.g.~vacuum stability, has been extensively investigated \citep{Degrassi:2012ry,Holthausen:2011aa,Iacobellis:2016eof,Haba:2013lga,Eichhorn:2014qka,Khan:2014kba,Khan:2015ipa,Helmboldt:2016mpi,PhysRevD.96.055020,Khan:2016sxm,Dev:2014yca,Dev:2017org}.

Another simple way to extend the SM is to add a second Higgs doublet. Supersymmetry is a common motivation for this addition, but supersymmetric models often require fine-tuning of parameters or considerable complication in order to predict a Higgs mass compatible with the combined ATLAS and CMS value of $m_h = 125.09 \pm 0.23\,$GeV \citep{Aad:2015zhl,Buttazzo:2013uya,Craig:2013cxa}. In general, the Two Higgs Doublet Model (2HDM) must account for the very SM-like nature of the Higgs \citep{Aad:2015zhl,Khachatryan:2014kca,Khachatryan:2014jba,Aad:2015gba} while  evading strong experimental bounds on its interactions.

The aim of this work is to consider whether the inclusion of an extra Higgs-doublet is compatible with both the existence of particular boundary conditions at the Planck scale and current theoretical and experimental constraints. In Section~\ref{sec:models} we will describe our two considered models, the Type-II 2HDM and the Inert Doublet Model (IDM). In Section~\ref{sec:THDM_numerical_analysis_constraints} we will then describe our methodology, including the theoretical and experimental constraints we apply to our scenarios. We will present our results for both models when confronted with boundary conditions for each of the Multiple Point Principle (MMP) or Asymptotic Safety in Section~\ref{sec:results}. We will find that neither model can accommodate the high scale boundary conditions of the MPP, while Asymptotic Safety remains viable. We will draw our conclusions in Section~\ref{sec:conclusions}. Finally, in Appendix~\ref{Appendix:RGEs} we will include the Renormalisation Group Equations (RGEs) of the the Higgs quartic-couplings for the reader's convenience.


\section{Considered Models}
\label{sec:models}

In this study we will focus on the Type-II 2HDM and the IDM, and present a brief summary of the models here in order to fix our notations and conventions. For useful reviews of these models see Refs.~\cite{Branco:2011iw} and \cite{Ilnicka:2015jba} respectively.

\subsection{The Two Higgs Doublet Model}
\label{sec:THDM}
The most general potential of the 2HDM is,
\begin{eqnarray}
	V \left(H_1, H_2 \right) &&= m_{11}^2 H_{1}^{\dagger} H_{1} + m_{22}^2 H_{2}^{\dagger} H_{2} - \left( m_{12}^2 H_{1}^{\dagger} H_{2} + c.c \right)  + \lambda_1 \left( H_{1}^{\dagger} H_{1} \right)^2 \\ \nonumber
	&&+ \lambda_2 \left( H_{2}^{\dagger} H_{2} \right)^2 + \lambda_3 \left( H_{1}^{\dagger} H_{1} \right) \left( H_{2}^{\dagger} H_{2} \right) + \lambda_4 \left( H_{1}^{\dagger} H_{2} \right) \left( H_{2}^{\dagger} H_{1} \right) \\ \nonumber
	&&+ \left( \frac{\lambda_5}{2} \left( H_{1}^{\dagger} H_{2} \right)^2 + \lambda_6 \left( H_{1}^{\dagger} H_{1} \right) \left( H_{1}^{\dagger} H_{2} \right) + \lambda_7 \left( H_{2}^{\dagger} H_{2} \right) \left( H_{1}^{\dagger} H_{2} \right) + c.c \right),
	\label{eq:THDM_potential}
\end{eqnarray}
where the two Higgs-doublets themselves are given by,
\begin{equation}
	H_n = \begin{pmatrix} \chi_n^{+} \\ \left( H_n^0 + i A_n^0 \right)/\sqrt{2} \end{pmatrix}, \quad n=1,2.
	\label{eq:THDM_Hn}
\end{equation}

The parameters $m_{11}^2$, $m_{22}^2$ and $\lambda_{1,2,3,4}$ are real, whilst $m_{12}^2$ and $\lambda_{5,6,7}$ can in principle be complex and induce CP violation. During electroweak symmetry breaking the neutral components of the Higgs fields, $H_n^0$, develop vacuum expectation values (vevs) $\langle H_n^0 \rangle = v_n / \sqrt{2}$. The expression $v = \sqrt{v_1^2 + v_2^2}$ is set to the SM Higgs vev's value of $246\,$GeV, but the ratio of the vevs, $\tan \beta = v_2/v_1$, is a free parameter. The physical scalar sector of the model includes two neutral scalar Higgs $h$ and $H$, a pseudoscalar Higgs $A$ and the charged Higgs $H^{\pm}$.

It's clear that the 2HDM potential is considerably more complicated than its Standard Model counterpart, so it's common to employ additional global symmetries to increase the predictivity of the model. There are only six possible types of global symmetry that have a distinctive effect on the potential \citep{Ivanov:2006yq,Ferreira:2015rha}. In this work we implement a $\mathbb{Z}_2$ symmetry to forbid Flavour Changing Neutral Currents (FCNCs) by allowing only one type of fermion to couple to one Higgs doublet. This requirement sets $\lambda_6$, $\lambda_7$ and $m_{12}^2$ to zero. However we then softly break this $\mathbb{Z}_2$ by re-introducing a positive non-zero $m_{12}^2$. For the results reported here we will restrict ourselves to a Type-II model where up-type quarks and leptons couple to the first Higgs-doublet and down-type quarks to the second Higgs-doublet. The dominant effect of the Yukawa sector on the running of the relevant Higgs parameters arises from the top-quark coupling, so we expect our results to be similar for for other 2HDM types. We checked this by repeating the analysis for the Type-I and flipped 2HDMs and found no significant differences from the results presented here.

For each parameter point the model is described by the parameters $m_{11}^2$ and $m_{22}^2$, which are replaced by $M_Z$ and $\tan \beta$ by applying the electroweak vacuum minimisation conditions, as well as the additional input parameters, $m_{12}^2$ and $\lambda_{i}(M_{\rm Pl})$ with $i=1\ldots 5$. We also use the top pole mass $m_t$ and the strong coupling constant $\alpha_S (M_Z)$ as inputs, allowing them to vary between $\pm 3 \sigma$ of their central values to account for the effect of their uncertainty on our results. Since we are interested in both the high and low scale behaviour of the potential's parameters we use SARAH 4.12.2 \citep{Staub:2013tta} to calculate the two-loop $\beta$ functions, which are used by FlexibleSUSY 2.0.1 \citep{Athron:2014yba,Athron:2017fvs,Allanach:2001kg,Allanach:2013kza} to run the couplings between $M_Z$ and $M_{\rm Pl}$.


\subsection{The Inert Doublet Model}
\label{sec:Inert}
We also consider the model where we introduce an additional unbroken $\mathbb{Z}_2$ symmetry, under which the new Higgs Doublet has odd parity but all other fields are even. The scalar sector now consists of the SM Higgs field $H$ and an inert doublet $\Phi$, with mixing between the two forbidden by the new symmetry. The inert doublet does not couple to any of the SM fields and does not gain a vacuum expectation value. The potential is,
\begin{eqnarray}
	V \left(H, \Phi \right) &&= m_{11}^2 H^{\dagger} H + m_{22}^2 \Phi^{\dagger} \Phi + \lambda_1 \left( H^{\dagger} H \right)^2 + \lambda_2 \left( \Phi^{\dagger} \Phi \right)^2 \\ \nonumber
	&&+ \lambda_3 \left( H^{\dagger} H \right) \left( \Phi^{\dagger} \Phi \right) + \lambda_4 \left( H^{\dagger} \Phi \right) \left( \Phi^{\dagger} H \right) + \left( \frac{\lambda_5}{2} \left( H^{\dagger} \Phi \right)^2 + h.c. \right),
	\label{eq:Inert_potential}
\end{eqnarray}
where all the parameters are real. Note that now the mixing term proportional to $m_{12}^2$ is absent. During electroweak symmetry breaking the neutral component of the SM Higgs doublet acquires a vacuum expectation value $v \approx 246\,$GeV. The neutral Higgs $h$ corresponds to the SM Higgs boson whilst $H$, $A$ and $H^{\pm}$ are inert scalars. The lightest of these $h_{LOP}$ (Lightest Odd Particle) is stable thanks to the $\mathbb{Z}_2$ symmetry and, assuming $h_{LOP}$ is one of the neutral scalars $H$ or $A$, it is a potential Dark Matter (DM) candidate \citep{PhysRevD.95.015017,PhysRevD.92.055006}.

The tree-level masses for the scalars are given by \citep{Goudelis:2013uca},
\begin{eqnarray}
m_h^2 &=& m_{11}^2 + 3 \lambda_1 v^2, \\ \nonumber
m_H^2 &=& m_{22}^2 + \frac{1}{2} \left(\lambda_3 + \lambda_4 + \lambda_5 \right) v^2, \\ \nonumber
m_A^2 &=& m_{22}^2 + \frac{1}{2} \left(\lambda_3 + \lambda_4 - \lambda_5 \right) v^2 ,\\ \nonumber
m_{H^{\pm}}^2 &=& m_{22}^2 + \frac{1}{2} \lambda_3 v^2.
\label{eq:Inert_scalar_tree_masses}
\end{eqnarray}
As in the previous case, we fix the mass term associated with the SM Higgs doublet $m_{11}^2$ via the electroweak minimisation conditions, but now don't have a second vev to fix $m_{22}^2$, which must remain an input. Our input parameters are therefore $m_{22}^2$ and $\lambda_{i}(M_{\rm Pl})$ with $i=1\ldots 5$. As in the Type-II model, we use SARAH and FlexibleSUSY to calculate the mass spectrum and to run couplings between the low and high scales of interest.


\section{Numerical Analysis and Constraints}
\label{sec:THDM_numerical_analysis_constraints}

\begin{table}[!tbp]
\centering
\begin{tabular}{|l|l|}
\hline
\multicolumn{2}{|c|}{Type-II Model Input} \\ \hline
$\lambda_{1,2} \left( M_{Pl} \right)$ & $0.0 - 1.0$         \\ \hline
$\lambda_{3,4} \left( M_{Pl} \right)$ & $-1.0 - 1.0$         \\ \hline
$\lambda_{5,6,7} \left( M_{Pl} \right)$ & $0.0$         \\ \hline
$m_{12}$ &   $0.0 - 2000$ $\mathrm{GeV}$ \\ \hline
$\tan{\beta}$ &   $2.0 - 50$ \\ \hline
\end{tabular} \qquad
\begin{tabular}{|l|l|}
\hline
\multicolumn{2}{|c|}{Inert Model Input} \\ \hline
$\lambda_{1,2} \left( M_{Pl} \right)$ & $0.0 - 1.0$         \\ \hline
$\lambda_{3,4} \left( M_{Pl} \right)$ & $-1.0 - 1.0$         \\ \hline
$\lambda_{5} \left( M_{Pl} \right)$ & $0.0$         \\ \hline
$m_{22}$ &   $0.0 - 2000$ $\mathrm{GeV}$ \\ \hline
\end{tabular}
\caption{Input parameter ranges for the numerical analysis of the \textbf{(left)} Type-II 2HDM and \textbf{(right)} IDM. Note that in the above, $m_{12}$ and $m_{22}$ are understood to be the square-roots of the input parameters $m_{12}^2$ and $m_{22}^2$ respectively.}
\label{tab:THDM_parameter_ranges}
\end{table}
The main focus of this work is the possibility and consequences of boundary conditions on all or some of the quartic couplings of the 2HDM and the IDM and their $\beta$ functions at the Planck scale,
\begin{equation}
\lambda_i \left( M_{Pl} \right), \beta_{\lambda_i} \left( M_{Pl} \right) = 0, \quad i = 1 \ldots 5
\label{eq:THDM_BCs}
\end{equation}
We use SARAH 4.12.2 \citep{Staub:2013tta} to calculate all of the model parameters, including mass matrices, tadpole equations, vertices and loop corrections, as well as the two-loop $\beta$ functions for each model. FlexibleSUSY 2.0.1 \citep{Athron:2014yba,Athron:2017fvs,Allanach:2001kg,Allanach:2013kza} uses this output to calculate the mass spectrum and to run the couplings between $M_{Z}$ and The Planck scale. Table \ref{tab:THDM_parameter_ranges} shows the input parameter ranges used in our scans for both the Type-II and Inert models.

Valid points in our parameter space scan are required to be perturbative up to the Planck scale. For the Higgs quartic couplings this requires them to satisfy $\lambda_i < \sqrt{4 \pi}$ up to $M_{Pl}$. We require the potential to be bounded-from-below at all scales up to $M_{Pl}$ \citep{Chowdhury:2015yja}. To that end we check if the conditions \citep{Branco:2011iw},
\begin{eqnarray}
\lambda_1 &>& 0, \\ \nonumber
\lambda_2 &>& 0, \\ \nonumber
\lambda_3 &>& -2 \sqrt{\lambda_1 \lambda_2}, \\ \nonumber
\lambda_3 + \lambda_4 - |\lambda_5| &>& -2 \sqrt{\lambda_1 \lambda_2},
\label{eq:THDM_VSCs}
\end{eqnarray}
are met at all scales \citep{Sher:1988mj,Chataignier:2018aud}. We use Vevacious \citep{Camargo-Molina:2013qva} to check if the EWSB minimum is the global minimum. Additionally, we require valid points to provide a SM Higgs candidate $124.7 \leq m_h \leq 127.1\,$GeV, where the allowed mass range is larger than the experimental error to additionally account for theoretical uncertainties.

Our aim is to find parameter choices that are compatible with perturbativity, vacuum stability and the SM Higgs mass, as well as other constraints on the Higgs boson from LHC Run-I, LEP and the Tevatron. We use 2HDMC 1.7.0 \citep{Eriksson:2009ws} to calculate  the relevant branching ratios required by HiggsBounds 4.3.1 \citep{Bechtle:2013wla} to apply 95\% confidence exclusions. This same input is also used by HiggsSignals 1.4.0 \citep{Bechtle:2013xfa} to perform a $\chi^2$ fit to the observed SM signal at the LHC \footnote{We note that new beta-versions of HiggsBounds-5 and HiggsSignals-2 that include $13\,$TeV LHC data were made available after this analysis was completed.}

In the case of the IDM we also apply constraints from analyses of LEP data \citep{Goudelis:2013uca}. Potential invisible decays of the $W$ and $Z$ boson via $W^\pm \to AH^\pm$, $W^\pm \to HH^\pm$, $Z \to AH$ and $Z\to H^+H^-$ are ruled out by the precise measurement of the $W$ and $Z$ boson widths. To prevent these, we require \citep{Gustafsson:2007pc,Cao:2007rm}
\begin{equation}
\text{Min}(M_A,M_H) +M_{H^\pm}  > M_W, \quad M_A + M_H > M_Z  \quad \text{and} \quad 2M_{H^\pm} > M_Z.
\label{eq:THDM_inert_collider_constraint_MZ}
\end{equation}
LEP constraints from searches for charginos and neutralinos \cite{Pierce:2007ut,Lundstrom:2008ai} are applied
by excluding the region where $M_A<100\,$GeV, $M_H<80\,$GeV and $M_A-M_H > 8\,$GeV simultaneously.
To ensure that our lightest odd particle is a neutral DM candidate we also insist on the following relation between the dark sector particles,
\begin{equation}
M_{H^{\pm}} > \text{min} \left( M_H, M_A \right).
\label{eq:THDM_inert_collider_constraint_LOP}
\end{equation}
We also look at constraints from electroweak precision observables for both of our models. The $S$, $T$ and $U$ parameters are calculated using 2HDMC and the results are checked against the current PDG limits \citep{Patrignani:2016xqp}, where
we require these precision observables within the range of $\pm 3 \sigma$. However, we note that these constraints do not restrict the parameter space beyond the bounds arising from the LHC Run-I, LEP and Tevatron described above.

In the 2HDM the existence of the charged Higgs bosons $H^{\pm}$ can affect the calculation of flavour observables. To take this into account we use SuperIso \citep{Mahmoudi:2007vz,Mahmoudi:2008tp,Mahmoudi:2009zz} to calculate the radiative $B$ meson decay $B \rightarrow X_s \gamma$, the leptonic $B$ decays $B_{s}^0 \rightarrow \mu^+ \mu^-$, $B_{d}^0 \rightarrow \mu^{+} \mu^{-}$, and $B \rightarrow \tau \nu$, the leptonic D decays $D \rightarrow \mu \nu$, $D_s \rightarrow \mu \nu$ and $D_s \rightarrow \tau \nu$ as well as the semileptonic decay $B \rightarrow D \tau \nu$, the kaon decay $K \rightarrow \mu \nu$ and the pion decay $\pi \rightarrow \mu \nu$. We then apply 95\% confidence level constraints on the branching ratios of these decays.

For the IDM, we use micrOMEGAS \citep{Belanger:2014vza} to calculate the DM relic density $\Omega h^2$, using the lightest of the neutral scalars $H$ and $A$ as the stable DM candidate. We compare the result to the combined experimental result from the WMAP \citep{Hinshaw:2012aka} and Planck \citep{Ade:2015xua} experiments,
\begin{equation}
\Omega h^2 = 0.1199 \pm 0.0027.
\label{eq:IDM_relic_density}
\end{equation}
We pass points that give a value less than $\Omega h^2 + 3 \sigma$ to allow for the possibility that the scalar DM candidate is not the only contribution to the relic density.

DM direct detection experiments place constraints on the spin independent scattering cross-section of Weakly Interacting Massive Particles (WIMPs) on nucleons. The strongest of these comes from the LUX \cite{Akerib:2016vxi} and XENON1T \citep{Aprile:2018dbl} experiments, which give constraints that are dependent on the mass of the WIMP DM candidate. We use micrOMEGAS to calculate the scattering cross sections for each of the points in our scan and exclude those that give values greater than the XENON1T constraints.


\begin{figure}[t!]
  \centering
	\subfloat[]{\includegraphics[width=0.5\textwidth]{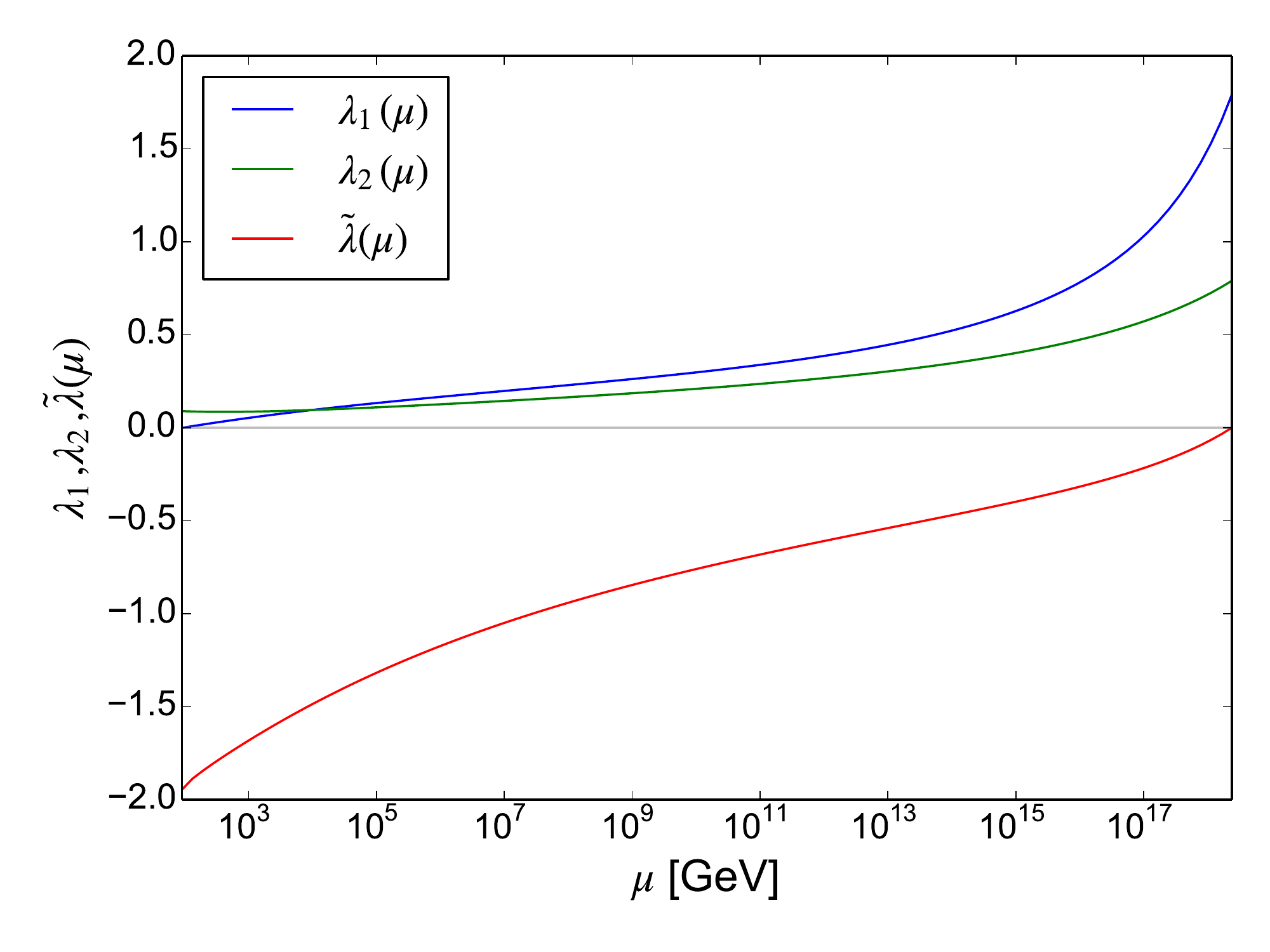}\label{fig:THDMII_MPP_vsc}}
  \hfill
  \subfloat[]{\includegraphics[width=0.5\textwidth]{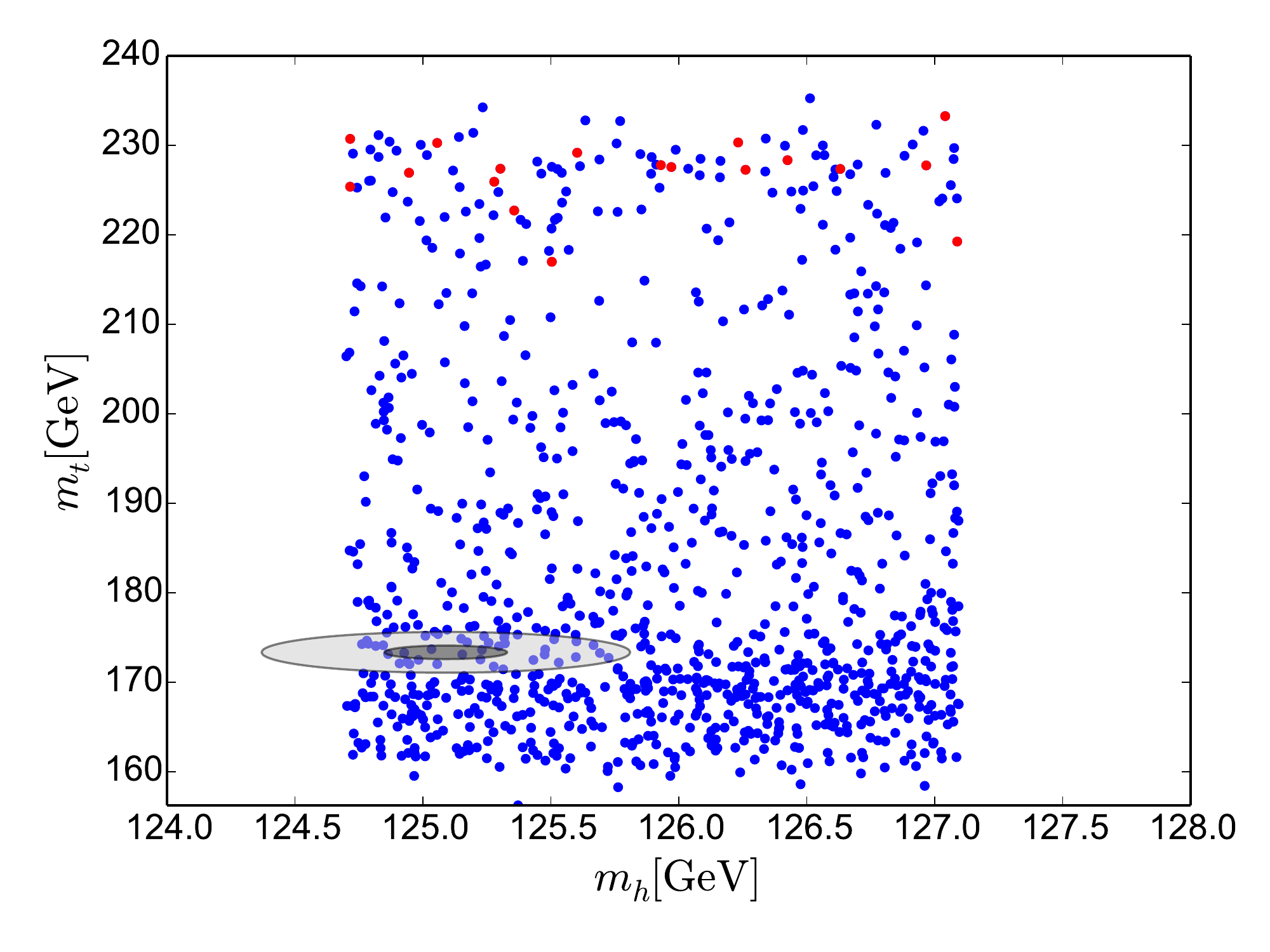}\label{fig:THDMII_MPP_mh_mt_comparison}}
\caption{\textbf{(a)} Example running of $\lambda_1$, $\lambda_2$ and $\tilde{\lambda}$ for a point that provides valid masses for the SM Higgs and the top quark in the Type-II 2HDM. Boundedness from below and vacuum stability requires that all three couplings are positive at all scales. \textbf{(b)} Results of our MPP scan in the $m_{h} - m_{t}$ plane of the Type-II 2HDM. The blue points provide valid SM Higgs masses whilst the red points also pass the vacuum stability conditions at all scales. The ellipses show the experimentally allowed values of $m_t$ and $m_{h}$ at $1 \,\sigma$ (dark grey) and $3 \,\sigma$ (light grey) uncertainty.}
\label{fig:THDMII_MPP_comparison}
\end{figure}

\section{Results}
\label{sec:results}

\subsection{The Multiple Point Principle in the Type-II Two Higgs Doublet Model}
\label{sec:THDM_MPP}
There are a number of possible scenarios that may enforce particular boundary conditions on the quartic Higgs couplings and their $\beta$ functions at the Planck scale \citep{Bhattacharyya:2017ksj}. One such scenario is the MPP \citep{Froggatt:1995rt} which posits that the effective potential has an additional minimum at the Planck scale, degenerate to the electroweak minimum. Applying the MPP in the SM leads to a prediction of the Higgs mass of $m_h = 129 \pm 1.5\,$GeV \citep{Buttazzo:2013uya}, which is not compatible with our current experimental value of $m_h$ but it is close enough to have inspired a number of investigations into the MPP in extensions of the SM \citep{Kawana:2014zxa,Haba:2016gqx,Hamada:2014xka,Nielsen:2017ows} and the 2HDM \citep{Froggatt:2004st,Froggatt:2006zc,Froggatt:2007qp}. The simplest scenario implementation of the MPP would be to have a global minimum at a high scale $\Lambda$, degenerate with the electroweak minimum, where all of the quartic couplings are zero at $\Lambda$, e.g $\lambda_i = 0, i = 1 \ldots 5$. However, the RGE running of $\lambda_1$ and $\lambda_2$ results in an unstable vacuum configuration \citep{Froggatt:2004st, Froggatt:2006zc,Froggatt:2007qp}.

It is possible for degenerate vacua to exist within the 2HDM if we relax the condition $\lambda_i = 0$. Specifically, by allowing $\lambda_1$, $\lambda_2$, $\lambda_3$ and $\lambda_4$ to be non-zero at $\Lambda$, the following conditions \citep{Froggatt:2004st} are consistent with the implementation of the MPP at $\Lambda$;
\begin{eqnarray}
\label{eq:THDM_MPP_conditions}
\lambda_5 \left( \Lambda \right) &=& 0 \\ \nonumber
\lambda_4 \left( \Lambda \right) &<& 0 \\ \nonumber
\tilde{\lambda} \left( \Lambda \right) = \sqrt{\lambda_1 \lambda_2} + \lambda_3 + \text{min}(0,\lambda_4) &=& 0 \\ \nonumber
\beta_{\tilde{\lambda}} \left( \Lambda \right) &=& 0,
\end{eqnarray}
where the form of $\tilde \lambda$ arises from the minimisation of the potential at $\Lambda$.
We note that setting these conditions at $\Lambda$ results in a potential with more symmetry than the original $\mathbb{Z}_2$ symmetry of \ref{eq:Inert_potential}.

To investigate whether these MPP conditions in the Type-II 2HDM are consistent with the current experimental constraints on the SM Higgs mass $m_h$ and the top pole mass $m_t$, we generated points in the parameter space as described in section \ref{sec:THDM_numerical_analysis_constraints}, applying the theoretical constraint of vacuum stability at all scales. Figure \ref{fig:THDMII_MPP_vsc} shows an example of the running of $\lambda_1$, $\lambda_2$ and $\tilde{\lambda}$ for a point that results in experimentally valid values of the SM Higgs mass and the top pole mass, and is also consistent with the MPP conditions \ref{eq:THDM_MPP_conditions}. Vacuum stability requires that all of these couplings remain greater than zero at all scales, but the running of $\tilde{\lambda}$ pulls it to negative values. Figure \ref{fig:THDMII_MPP_mh_mt_comparison} shows values for the SM-like Higgs mass and top-quark mass arising from the new MPP boundary conditions, where red points correspond to choices with a stable potential and blue points to those that violate the stability conditions. Although there are many blue points with acceptable Higgs and top-quark masses, there are no satisfactory red points. Parameter choices that satisfy the vacuum stability conditions (red) have larger values of the top Yukawa $y_t$ which positively contribute to the running of the quartic couplings. The larger required $y_t$ corresponds to a top mass in the range $220 \lesssim m_t \lesssim 230\,$GeV which is not compatible with current experimental bounds on the top pole mass.


\subsection{Asymptotic Safety in the Type-II Two Higgs Doublet Model}
\label{sec:THDM_AS}

Another possibilty for the high scale dynamics that enforces high scales boundary conditions is \textit{Asymptotic Safety}, in which the quartic couplings of the Higgs sector run towards an ultraviolet interacting fixed point \cite{Litim:2014uca,Litim:2015iea,Bond:2017wut,Bond:2016dvk,Sannino:2014lxa,Bajc:2016efj,Pelaggi:2017wzr,Pelaggi:2017abg,Ipek:2018sai}. It has been suggested that gravitational contributions may become significant at very high scales and alter the running of the couplings of the scalar potential to provide such a boundary condition \cite{Shaposhnikov:2009pv,Wetterich:2011aa,Eichhorn:2017sok,Eichhorn:2017lry,Eichhorn:2017ylw}. In the context of the 2HDM, we are therefore seeking scenarios that exhibit zero values for the $\beta$-functions of the Higgs quartic couplings at the Planck scale whilst allowing the couplings themselves to be non-zero.

Note that it is important at this stage to be clear on what we mean by a $\beta$-function being zero. For each of the points in our parameter space we perform a perturbative calculation of the RGE evolution of the model couplings, and accommodate the uncertainty associated with this calculation by allowing for small, non-zero values of the $\beta$-functions. We estimate this uncertainty by using the difference between the $\beta$-function values at $M_{Pl}$ calculated using one-loop and two-loop RGEs, and we consider the $\beta$-function to be zero if it is smaller that the RGE truncation error. In the case of the 2HDM we calculated this is,
\begin{eqnarray}
\label{eq:THDM_truncation_error}
\beta_{\lambda_1} \left( M_{Pl} \right) &<& 0.0127 \\ \nonumber
\beta_{\lambda_2} \left( M_{Pl} \right) &<& 0.0064 \\ \nonumber
\beta_{\lambda_3} \left( M_{Pl} \right) &<& 0.0139 \\ \nonumber
\beta_{\lambda_4} \left( M_{Pl} \right) &<& 0.0030.
\end{eqnarray}

\begin{figure}[t!]
  \centering
  \subfloat[]{\includegraphics[width=0.5\textwidth]{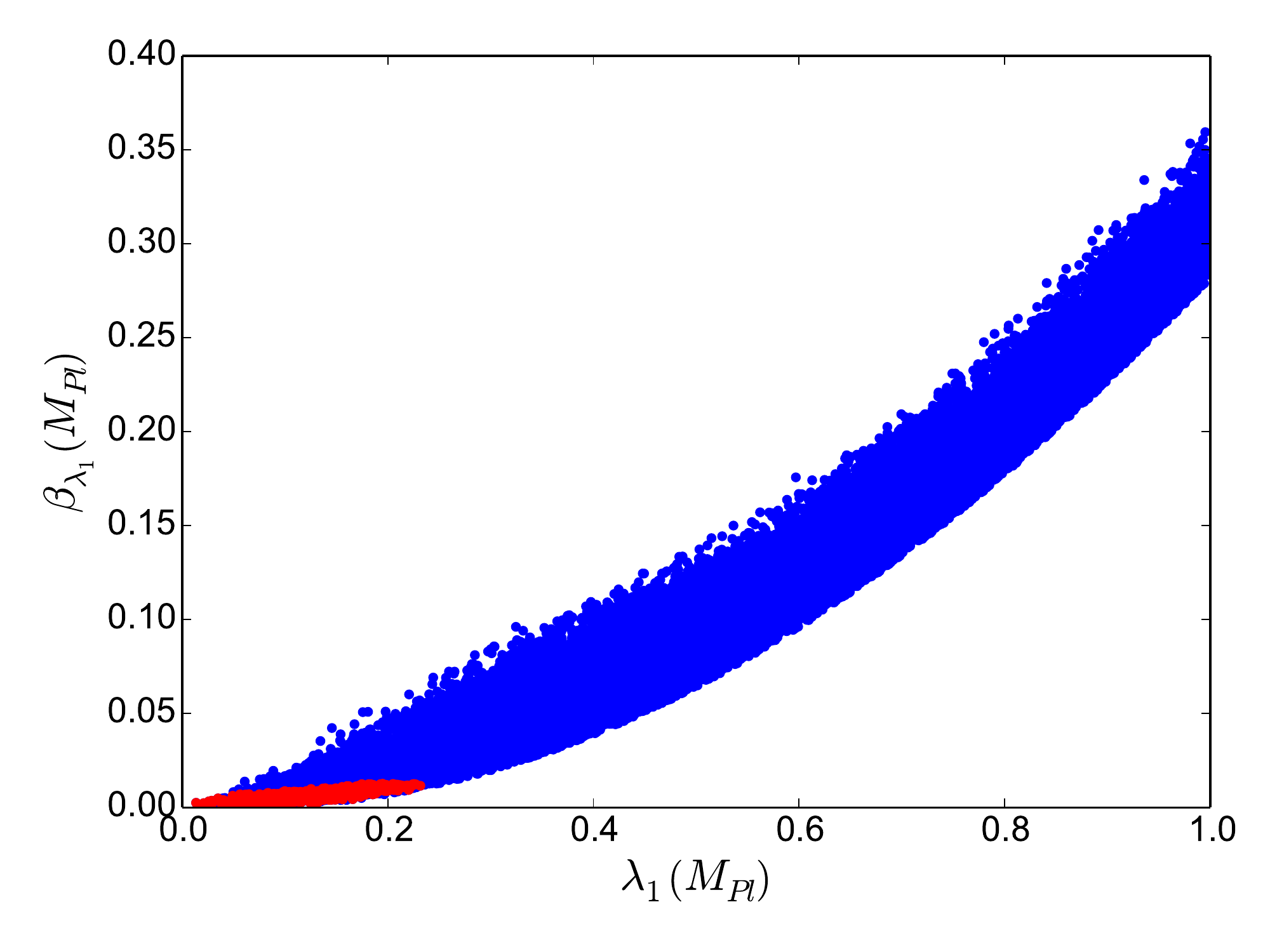}\label{fig:THDMII_l1_betal1_comparison_theoretical}}
  \hfill
  \subfloat[]{\includegraphics[width=0.5\textwidth]{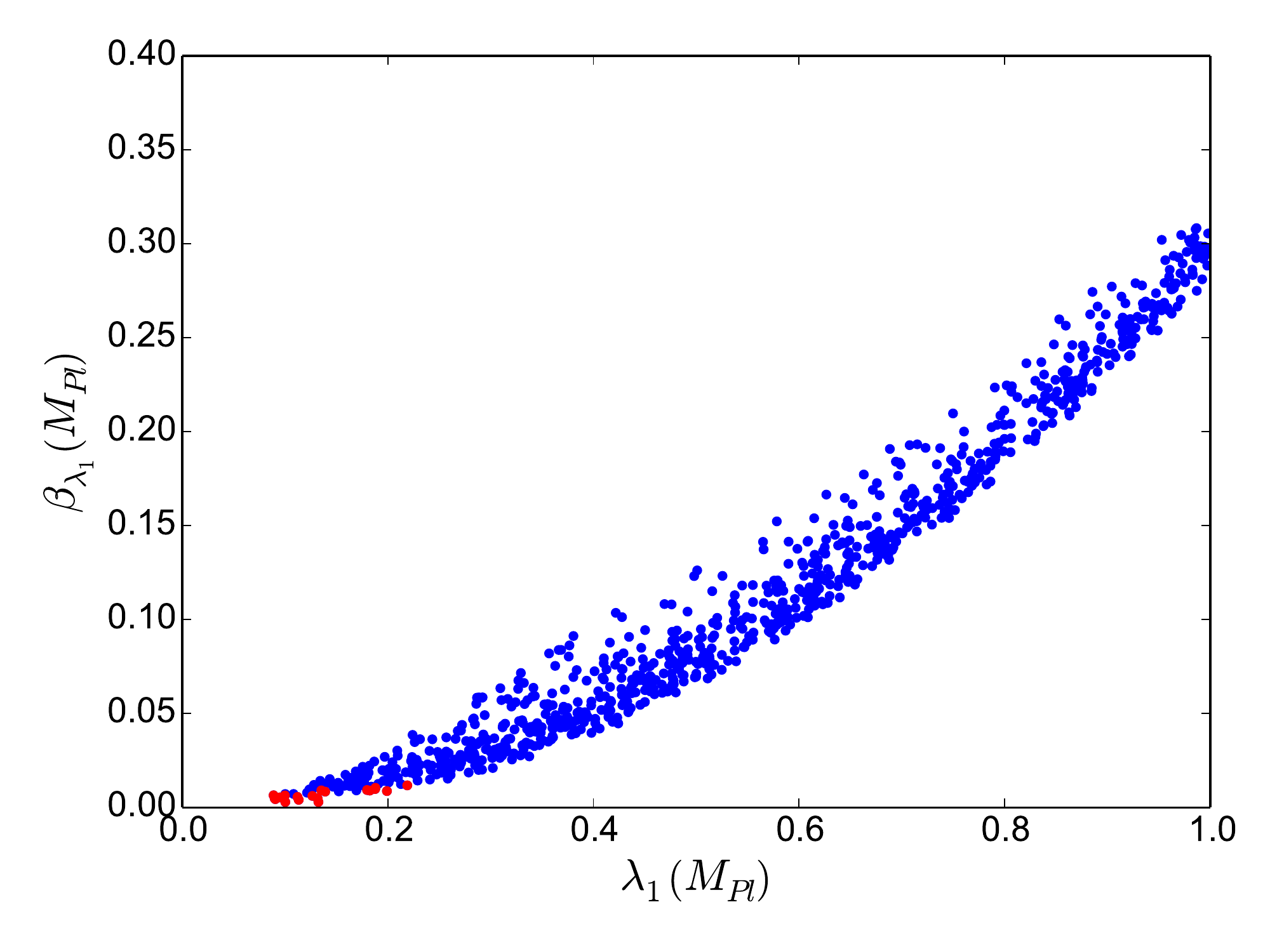}\label{fig:THDMII_l1_betal1_comparison_experimental}}
\caption{Compatible values of the Higgs quartic coupling $\lambda_1 \left( M_{\rm Pl} \right)$ against $\beta_{\lambda_1} \left( M_{\rm Pl} \right)$ in the Type II 2HDM. \textbf{(a)} includes points that are stable and perturbative up to $M_{Pl}$ and include an SM Higgs candidate, whilst \textbf{(b)} also enforces all relevant experimental constraints discussed in section \ref{sec:THDM_numerical_analysis_constraints}. Blue points obey $\beta_{\lambda_{1,2,3,4}} < 1.0$ at $M_{Pl}$ whilst red points obey $\beta_{\lambda_1} < 0.0127$, $\beta_{\lambda_2} < 0.0064$, $\beta_{\lambda_3} < 0.0139$, $\beta_{\lambda_4} < 0.0030$ at $M_{Pl}$.}
\label{fig:THDMII_l1_betal1_comparison}
\end{figure}

\begin{figure}[tbh]
  \centering
  \subfloat[]{\includegraphics[width=0.5\textwidth]{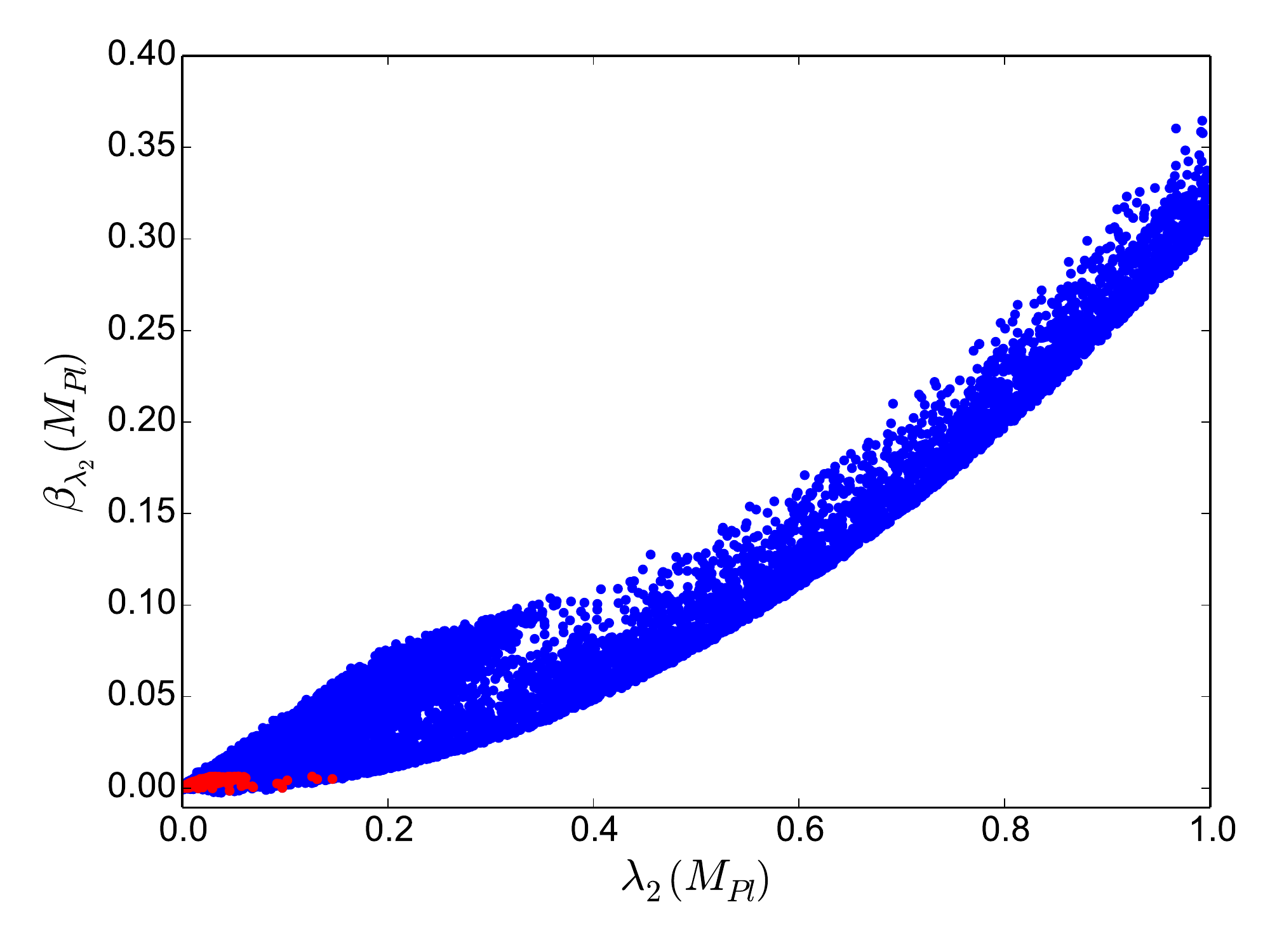}\label{fig:THDMII_l2_betal2_comparison_theoretical}}
  \hfill
  \subfloat[]{\includegraphics[width=0.5\textwidth]{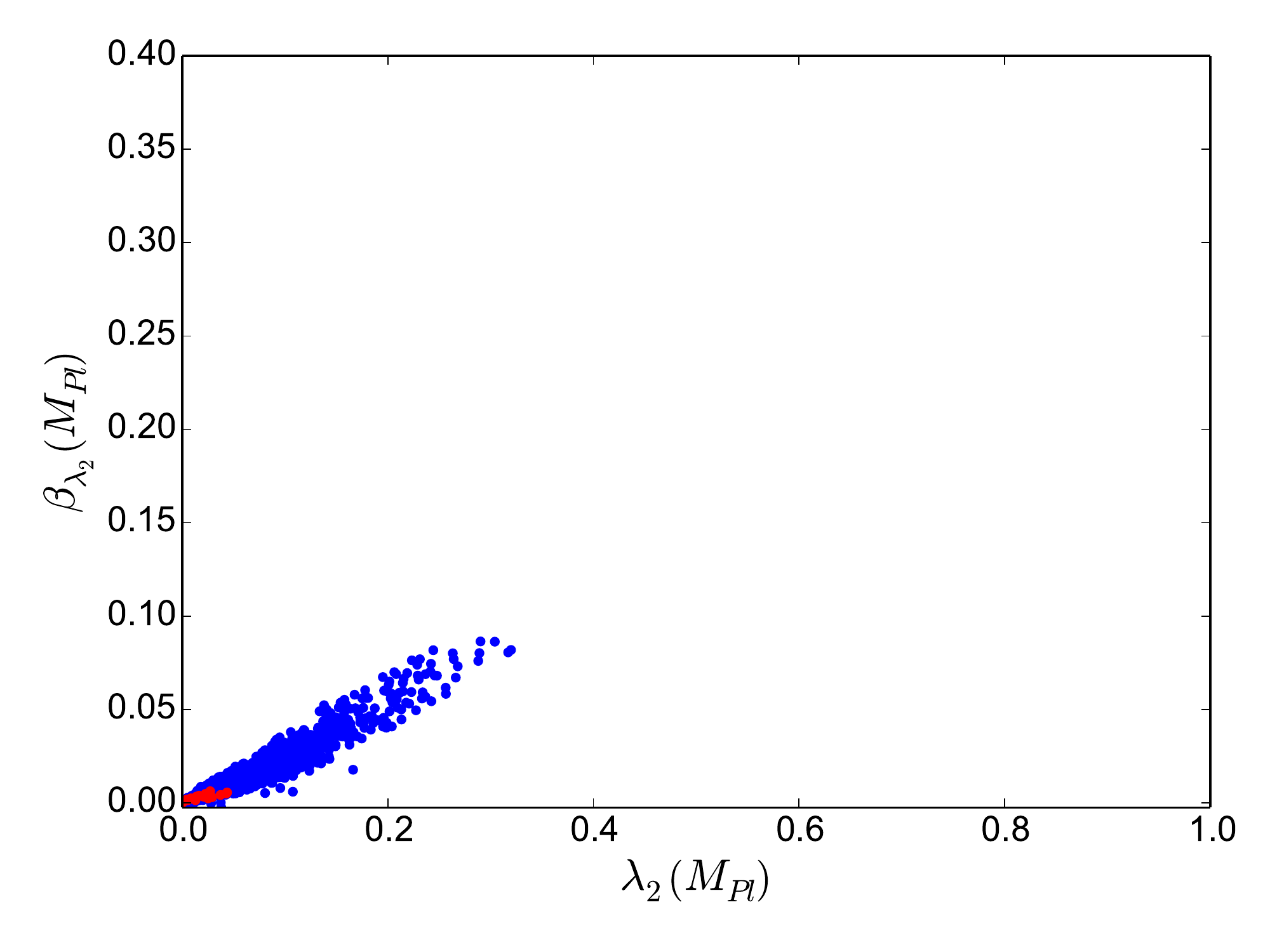}\label{fig:THDMII_l2_betal2_comparison_experimental}}
\caption{Compatible values of the Higgs quartic coupling $\lambda_2 \left( M_{\rm Pl} \right)$ against $\beta_{\lambda_2} \left( M_{\rm Pl} \right)$ in the Type II 2HDM. \textbf{(a)} includes points that are stable and perturbative up to $M_{Pl}$ and include an SM Higgs candidate, whilst \textbf{(b)} also enforces all relevant experimental constraints discussed in section \ref{sec:THDM_numerical_analysis_constraints}. Blue points obey $\beta_{\lambda_{1,2,3,4}} < 1.0$ at $M_{Pl}$ whilst red points obey $\beta_{\lambda_1} < 0.0127$, $\beta_{\lambda_2} < 0.0064$, $\beta_{\lambda_3} < 0.0139$, $\beta_{\lambda_4} < 0.0030$ at $M_{Pl}$. \\ \\}
\label{fig:THDMII_l2_betal2_comparison}
\end{figure}

\begin{figure}[tbh]
  \centering
  \subfloat[]{\includegraphics[width=0.5\textwidth]{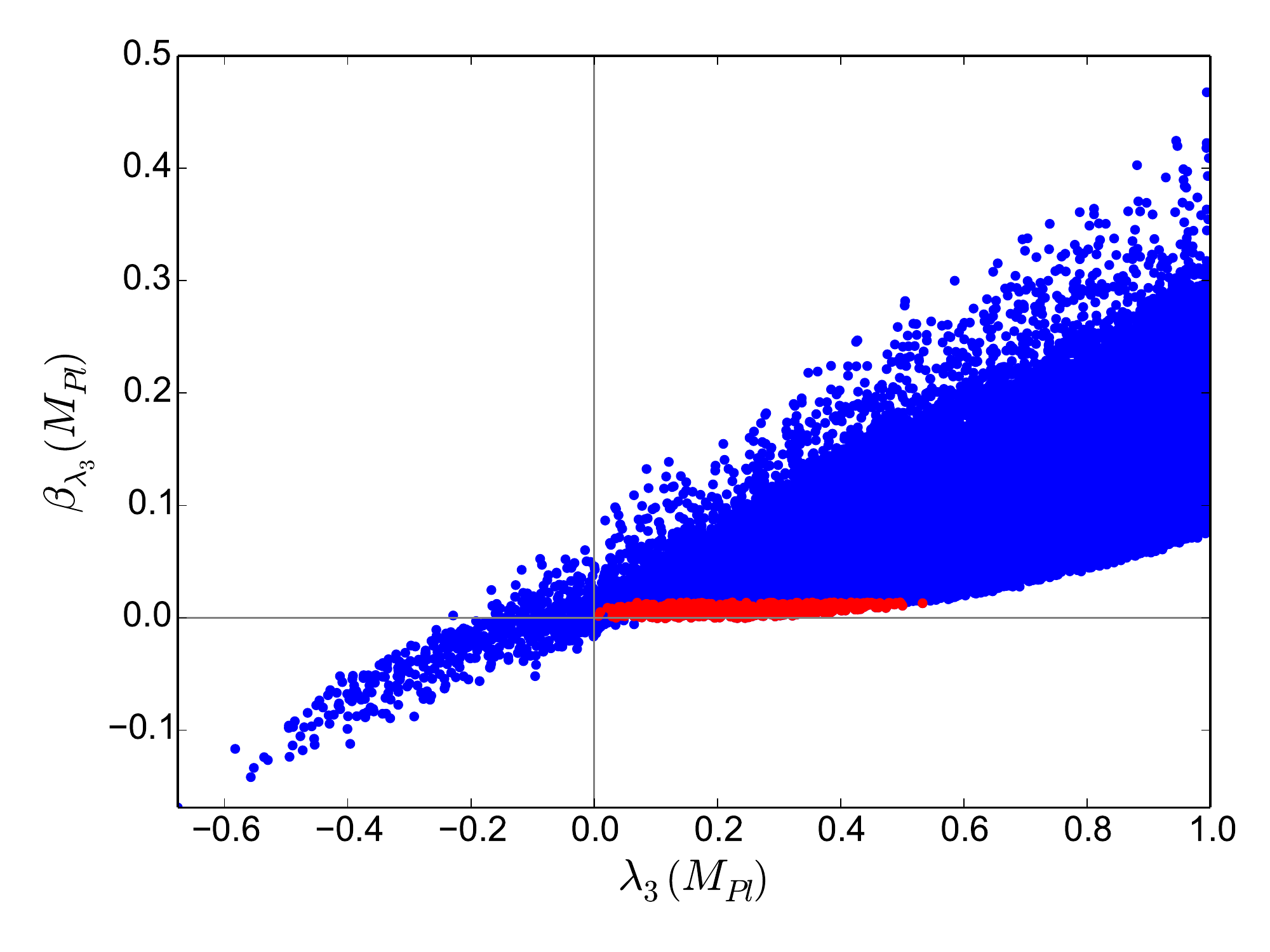}\label{fig:THDMII_l3_betal3_comparison_theoretical}}
  \hfill
  \subfloat[]{\includegraphics[width=0.5\textwidth]{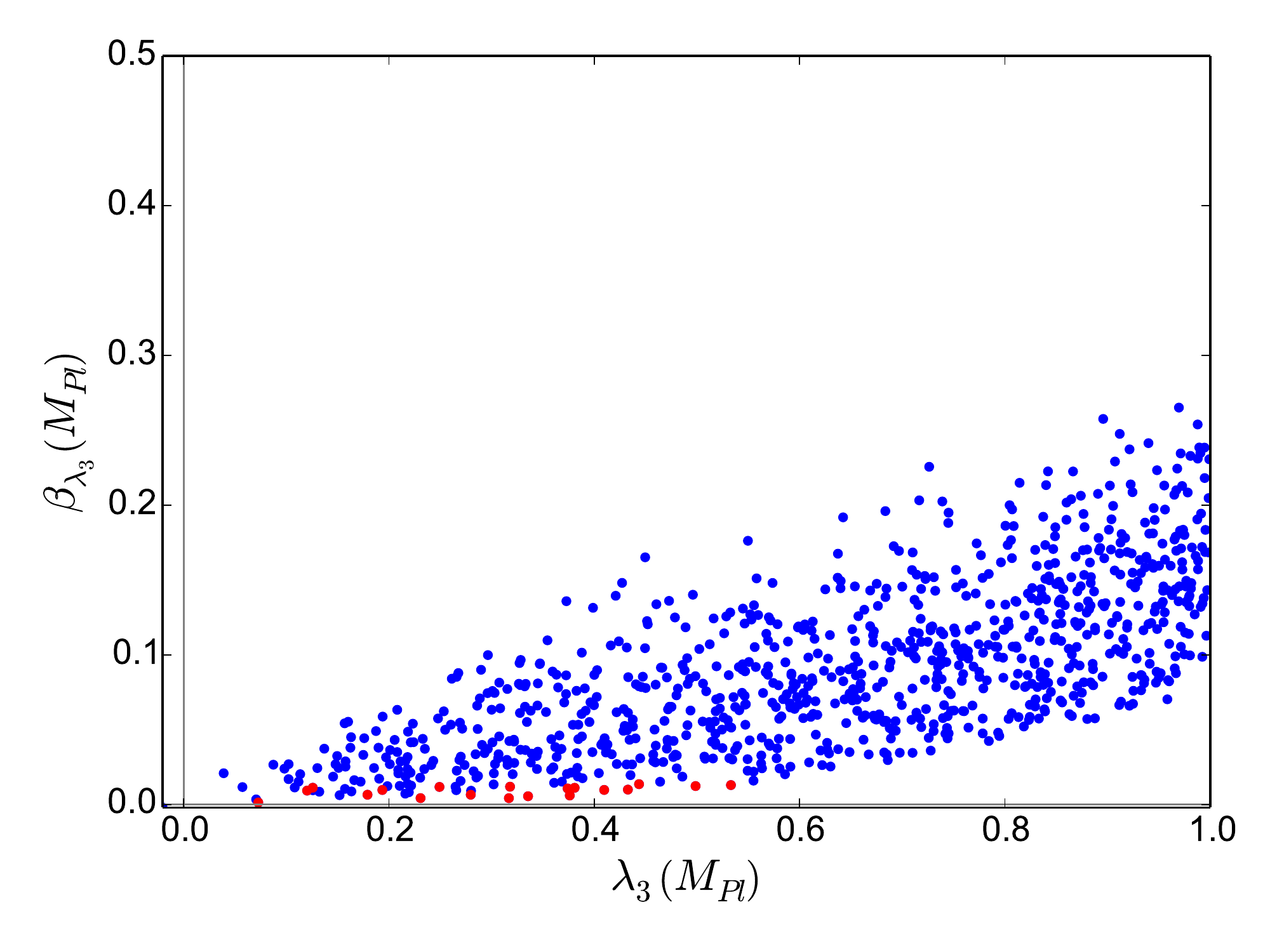}\label{fig:THDMII_l3_betal3_comparison_experimental}}
\caption{Compatible values of the Higgs quartic coupling $\lambda_3 \left( M_{\rm Pl} \right)$ against $\beta_{\lambda_3} \left( M_{\rm Pl} \right)$ in the Type II 2HDM. \textbf{(a)} includes points that are stable and perturbative up to $M_{Pl}$ and include an SM Higgs candidate, whilst \textbf{(b)} also enforces all relevant experimental constraints discussed in section \ref{sec:THDM_numerical_analysis_constraints}. Blue points obey $\beta_{\lambda_{1,2,3,4}} < 1.0$ at $M_{Pl}$ whilst red points obey $\beta_{\lambda_1} < 0.0127$, $\beta_{\lambda_2} < 0.0064$, $\beta_{\lambda_3} < 0.0139$, $\beta_{\lambda_4} < 0.0030$ at $M_{Pl}$. \\ \\}
\label{fig:THDMII_l3_betal3_comparison}
\end{figure}

\begin{figure}[t!]
  \centering
  \subfloat[]{\includegraphics[width=0.5\textwidth]{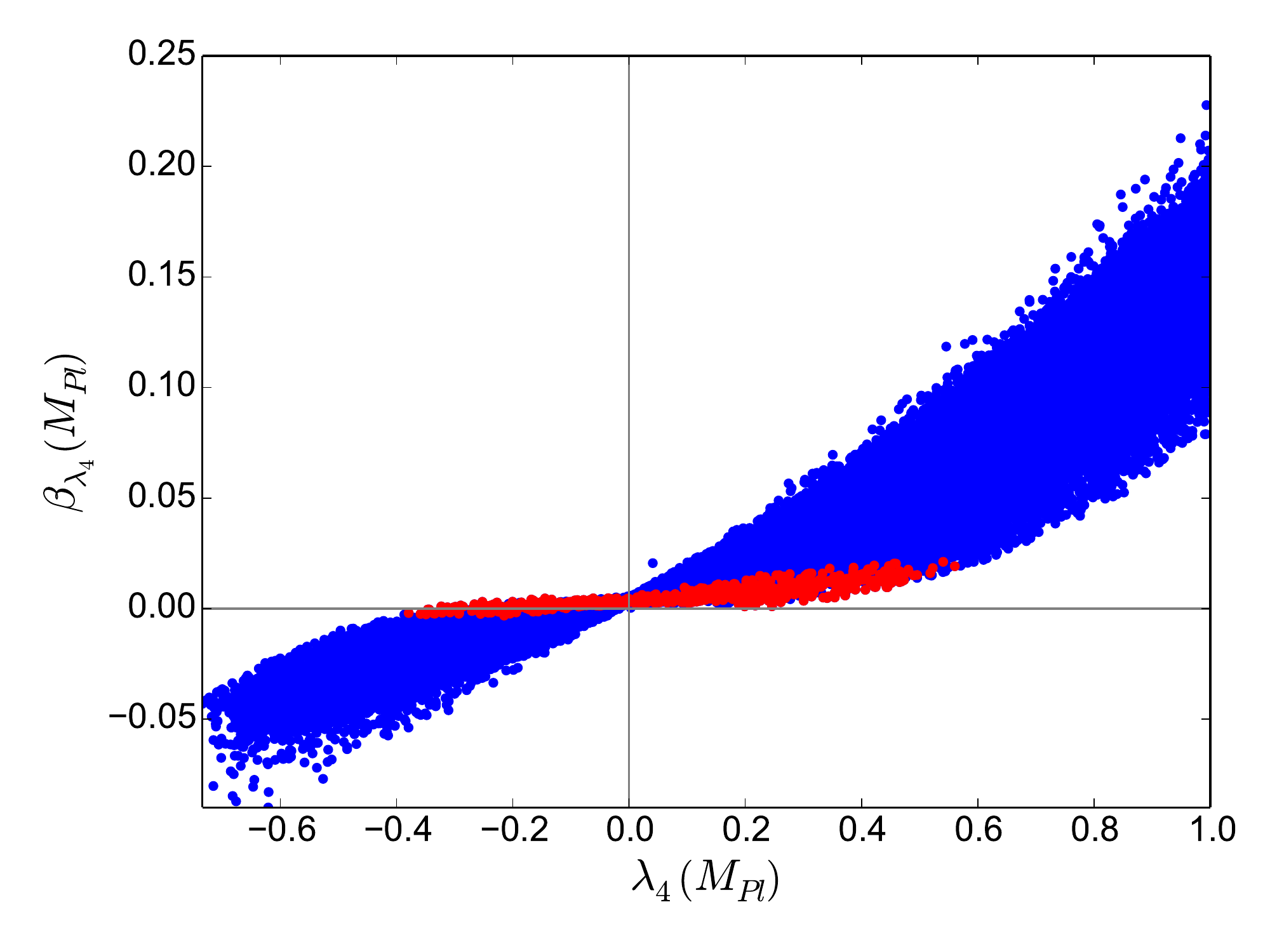}\label{fig:THDMII_l4_betal4_comparison_theoretical}}
  \hfill
  \subfloat[]{\includegraphics[width=0.5\textwidth]{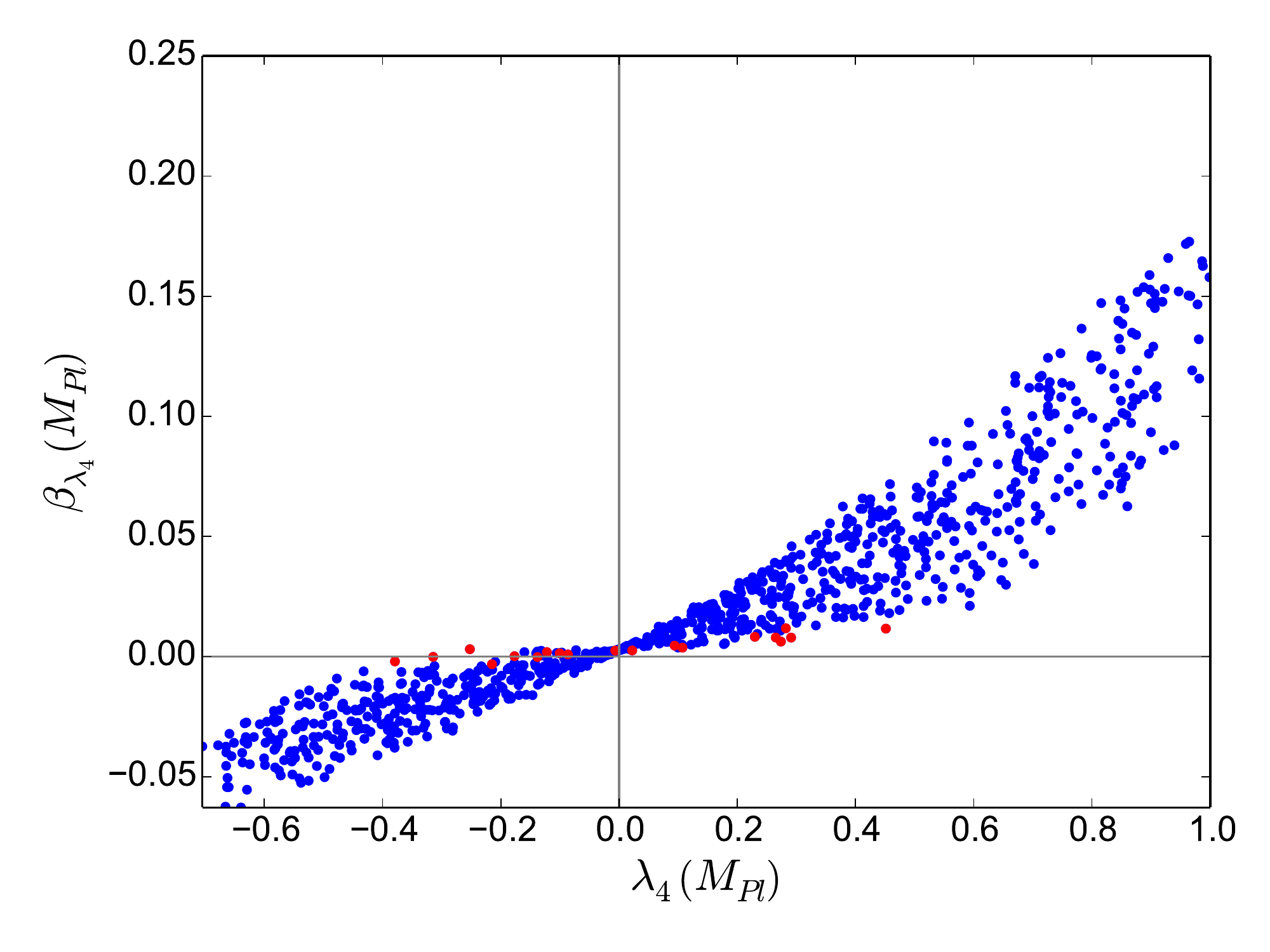}\label{fig:THDMII_l4_betal4_comparison_experimental}}
\caption{Compatible values of the Higgs quartic coupling $\lambda_4 \left( M_{\rm Pl} \right)$ against $\beta_{\lambda_4} \left( M_{\rm Pl} \right)$ in the Type II 2HDM. \textbf{(a)} includes points that are stable and perturbative up to $M_{Pl}$ and include an SM Higgs candidate, whilst \textbf{(b)} also enforces all relevant experimental constraints discussed in section \ref{sec:THDM_numerical_analysis_constraints}. Blue points obey $\beta_{\lambda_{1,2,3,4}} < 1.0$ at $M_{Pl}$ whilst red points obey $\beta_{\lambda_1} < 0.0127$, $\beta_{\lambda_2} < 0.0064$, $\beta_{\lambda_3} < 0.0139$, $\beta_{\lambda_4} < 0.0030$ at $M_{Pl}$.}
\label{fig:THDMII_l4_betal4_comparison}
\end{figure}

We now present the results of our numerical analysis of the Type-II 2HDM, in which we look for regions of parameter space that are compatible with the high scale boundary conditions that can arise under the requirement for asymptotic safety. We apply the relevant theoretical and experimental constraints described in Section \ref{sec:THDM_numerical_analysis_constraints} as well the $\beta_{\lambda_i} = 0$ constraints shown in Eq.~\ref{eq:THDM_truncation_error}. \Cref{fig:THDMII_l1_betal1_comparison,fig:THDMII_l2_betal2_comparison,fig:THDMII_l3_betal3_comparison,fig:THDMII_l4_betal4_comparison} show the values of the four non-zero quartic Higgs couplings $\lambda_{1,2,3,4}$ and their $\beta$ functions. The left plots include the theoretical constraints of perturbativity, vacuum stability and a valid SM Higgs candidate, whilst those on the right also include experimental constraints. Points in red provide values of the $\beta$ functions that are compatible with our asymptotic safety high scale boundary conditions, whilst those in blue do not pass those constraints. Clearly there are regions of parameter space where all of the $\beta$ functions of the quartic Higgs couplings are within the truncation errors, even after all of the relevant experimental constraints have been applied. These regions correspond to very small but non-zero values of the quartic couplings at $M_{Pl}$, consistent with a UV interacting fixed point.

\begin{figure}[t!]
  \centering
  \subfloat[]{\includegraphics[width=0.5\textwidth]{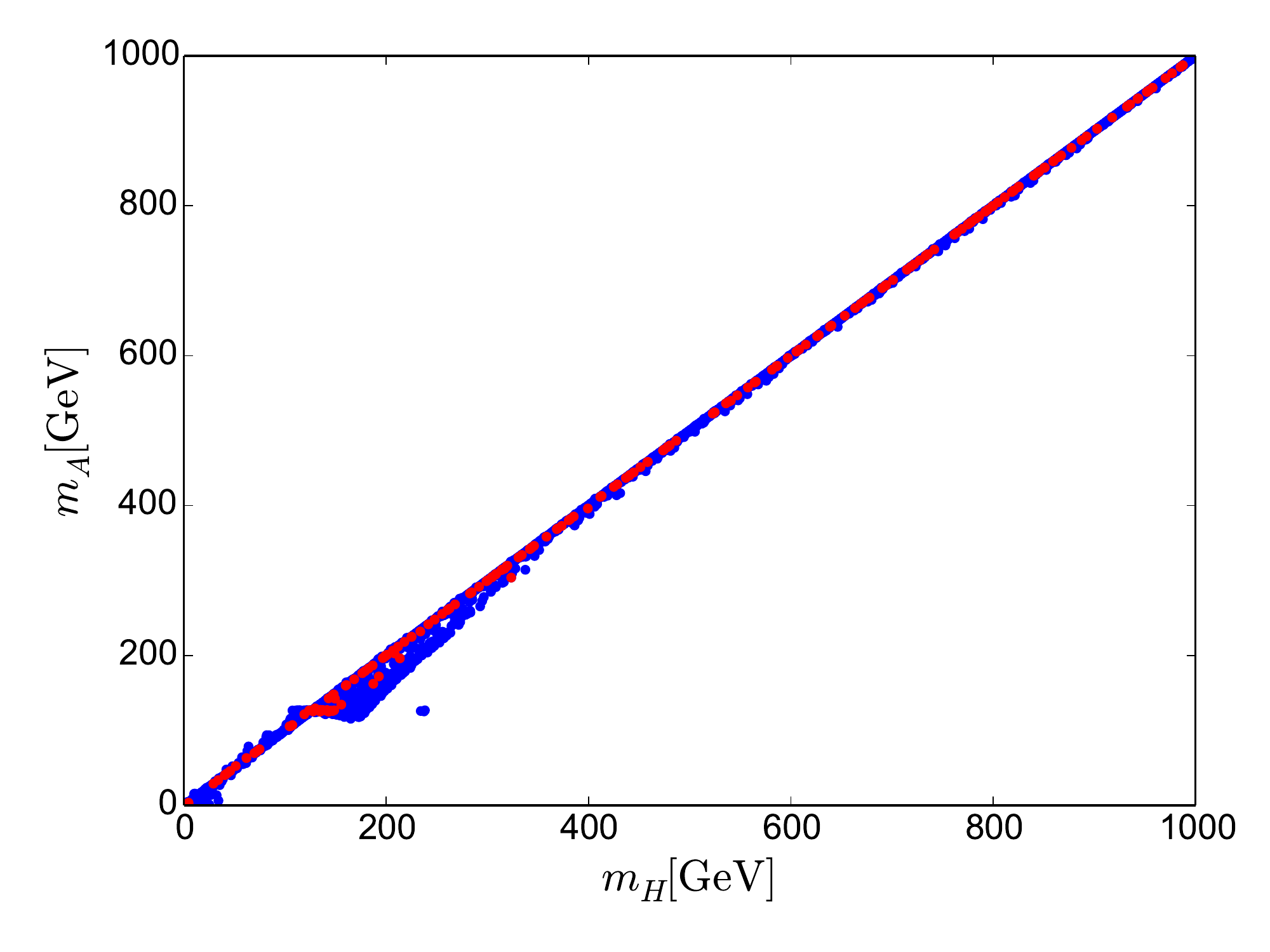}\label{fig:THDMII_heavyhiggs_A0higgs_comparison_theoretical}}
  \hfill
  \subfloat[]{\includegraphics[width=0.5\textwidth]{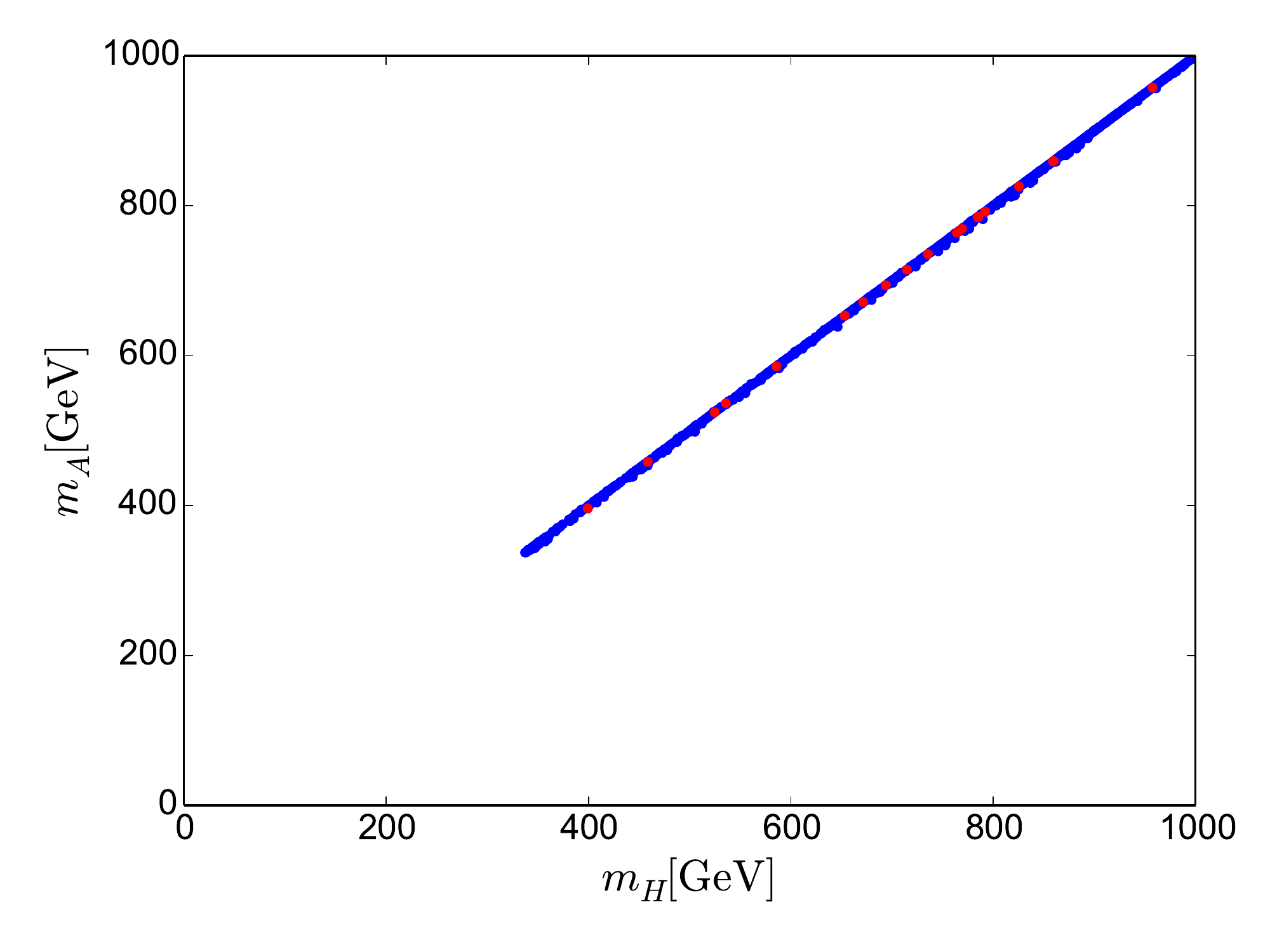}\label{fig:THDMII_heavyhiggs_A0higgs_comparison_experimental}}
\caption{Compatible values of the  heavy neutral Higgs mass $m_H$ against the pseudoscalar Higgs $m_A$ in the Type II 2HDM. \textbf{(a)} includes points that are stable and perturbative up to $M_{Pl}$ and include an SM Higgs candidate, whilst \textbf{(b)} also enforces all relevant experimental constraints discussed in section \ref{sec:THDM_numerical_analysis_constraints}. Blue points obey $\beta_{\lambda_{1,2,3,4}} < 1.0$ at $M_{Pl}$ whilst red points obey $\beta_{\lambda_1} < 0.0127$, $\beta_{\lambda_2} < 0.0064$, $\beta_{\lambda_3} < 0.0139$, $\beta_{\lambda_4} < 0.0030$ at $M_{Pl}$.}
\label{fig:THDMII_heavyhiggs_A0higgs_comparison}
\end{figure}

\begin{figure}[t!]
  \centering
  \subfloat[]{\includegraphics[width=0.5\textwidth]{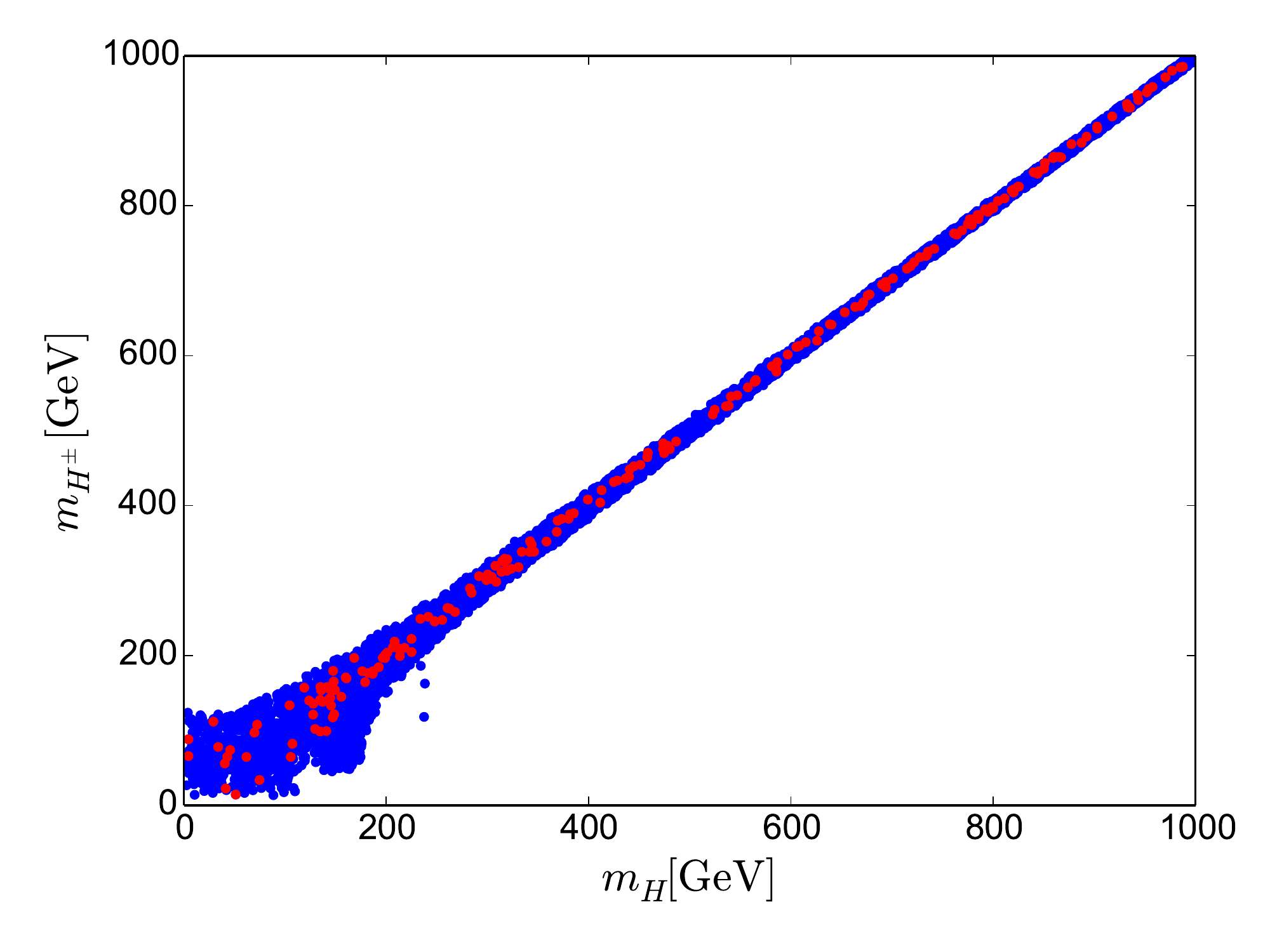}\label{fig:THDMII_heavyhiggs_chargedhiggs_comparison_theoretical}}
  \hfill
  \subfloat[]{\includegraphics[width=0.5\textwidth]{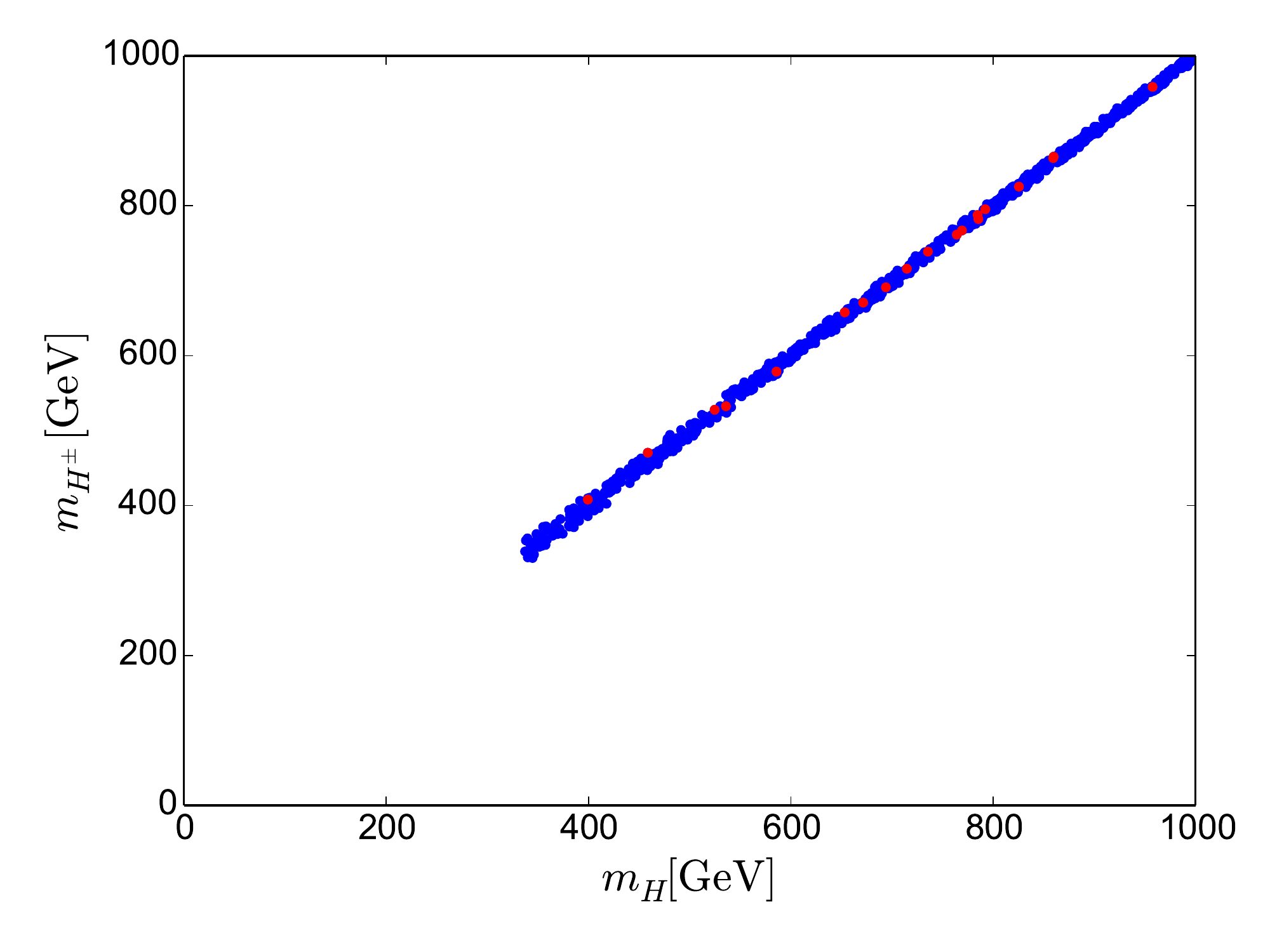}\label{fig:THDMII_heavyhiggs_chargedhiggs_comparison_experimental}}
\caption{Compatible values of the  heavy neutral Higgs mass $m_H$ against the charged Higgs $m_{H^{\pm}}$ in the Type II 2HDM. \textbf{(a)} includes points that are stable and perturbative up to $M_{Pl}$ and include an SM Higgs candidate, whilst \textbf{(b)} also enforces all relevant experimental constraints discussed in section \ref{sec:THDM_numerical_analysis_constraints}. Blue points obey $\beta_{\lambda_{1,2,3,4}} < 1.0$ at $M_{Pl}$ whilst red points obey $\beta_{\lambda_1} < 0.0127$, $\beta_{\lambda_2} < 0.0064$, $\beta_{\lambda_3} < 0.0139$, $\beta_{\lambda_4} < 0.0030$ at $M_{Pl}$. \\ \\}
\label{fig:THDMII_heavyhiggs_chargedhiggs_comparison}
\end{figure}

Figure \ref{fig:THDMII_heavyhiggs_A0higgs_comparison} shows the masses of the heavy neutral scalar $m_{H}$ against the pseudoscalar Higgs mass $m_{A}$, whilst Figure \ref{fig:THDMII_heavyhiggs_chargedhiggs_comparison} compares it with the charged Higgs mass $m_{H^{\pm}}$. As the scale associated with the the additional Higgs becomes significantly larger than the electroweak scale, the second doublet decouples from the first and the masses of $H$, $A$, and $H^{\pm}$ become degenerate. A lower limit on the masses of the extra scalars of around $m_{H,A,H^{\pm}} \approx 330\,$GeV is enforced once we apply the collider and flavour constraints. However, the points that are consistent with our high scale $\beta$ function conditions can have a range of different masses, and those conditions do not apply strong constraints upon the scalar mass spectrum in the Type-II 2HDM.

\subsection{The Multiple Point Principle in the Inert Doublet Model}
\label{sec:THDM_Inert_MPP}

Eq. \ref{eq:THDM_MPP_conditions} provides the  conditions that a 2HDM parameter point must satisfy to be consistent with the MPP. These constraints also apply to the IDM. We examined the IDM parameter space in the same way as we did for the Type-II 2HDM case detailed in Section \ref{sec:THDM_MPP}. We applied the MPP conditions at $M_{Pl}$ and required valid points to be stable up to the Planck scale and to have a SM Higgs candidate.

\begin{figure}[t!]
  \centering
\includegraphics[width=0.5\textwidth]{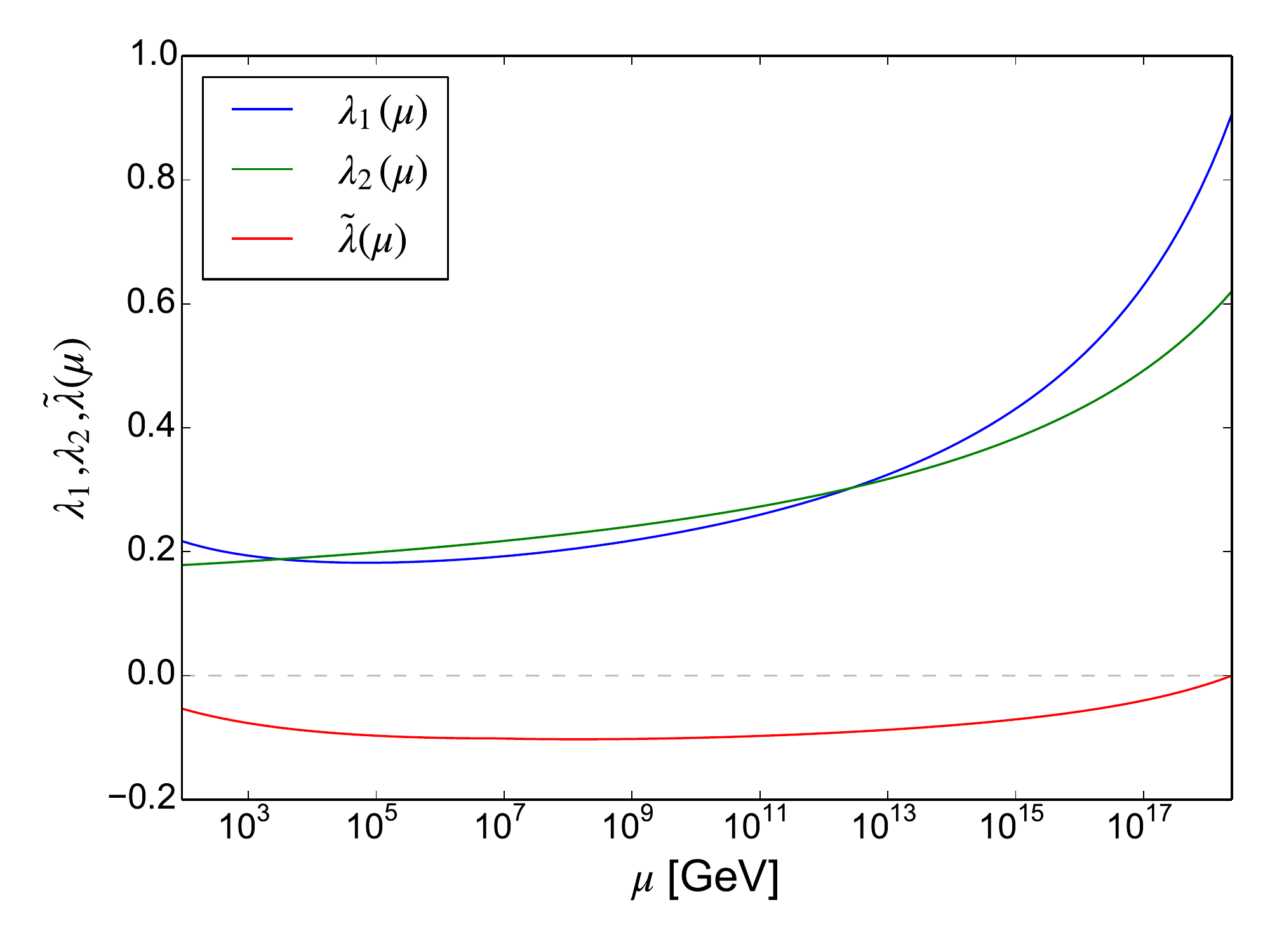}
\caption{Example running of $\lambda_1$, $\lambda_2$ and $\tilde{\lambda}$ for a point that provides valid masses for the SM Higgs and the top quark in the IDM. Boundedness from below and vacuum stability requires that all three couplings are positive at all scales.}
\label{fig:Inert_MPP_vsc}
\end{figure}

Figure \ref{fig:Inert_MPP_vsc} shows the running of the quartic couplings $\lambda_1$, $\lambda_2$ and  $\tilde{\lambda}$ for an example point in our scan that provided a valid SM Higgs and top mass. As in the Type-II model, a stable vacuum requires all three of these couplings to be positive at all scales. Clearly this point fails our vacuum stability test, and is representative of the other points in our scan. We found {\em no points} that could simultaneously satisfy the constraints of perturbativity, vacuum stability and the requirement of a realistic SM mass spectrum. Specifically, there are points that provide valid SM Higgs and top masses, but all of these points fail the condition $\tilde{\lambda} > 0$. In fact, we found no points that could satisfy the MPP conditions outlined in Eq. \ref{eq:THDM_MPP_conditions} that remained stable up to the Planck scale, regardless of their Higgs or top masses. This therefore suggests that the MPP cannot be implemented successfully in the IDM.

\subsection{Asymptotic Safety in the Inert Doublet Model}
\label{sec:THDM_Inert_AS}

\begin{figure}[tbh]
  \centering
  \subfloat[]{\includegraphics[width=0.5\textwidth]{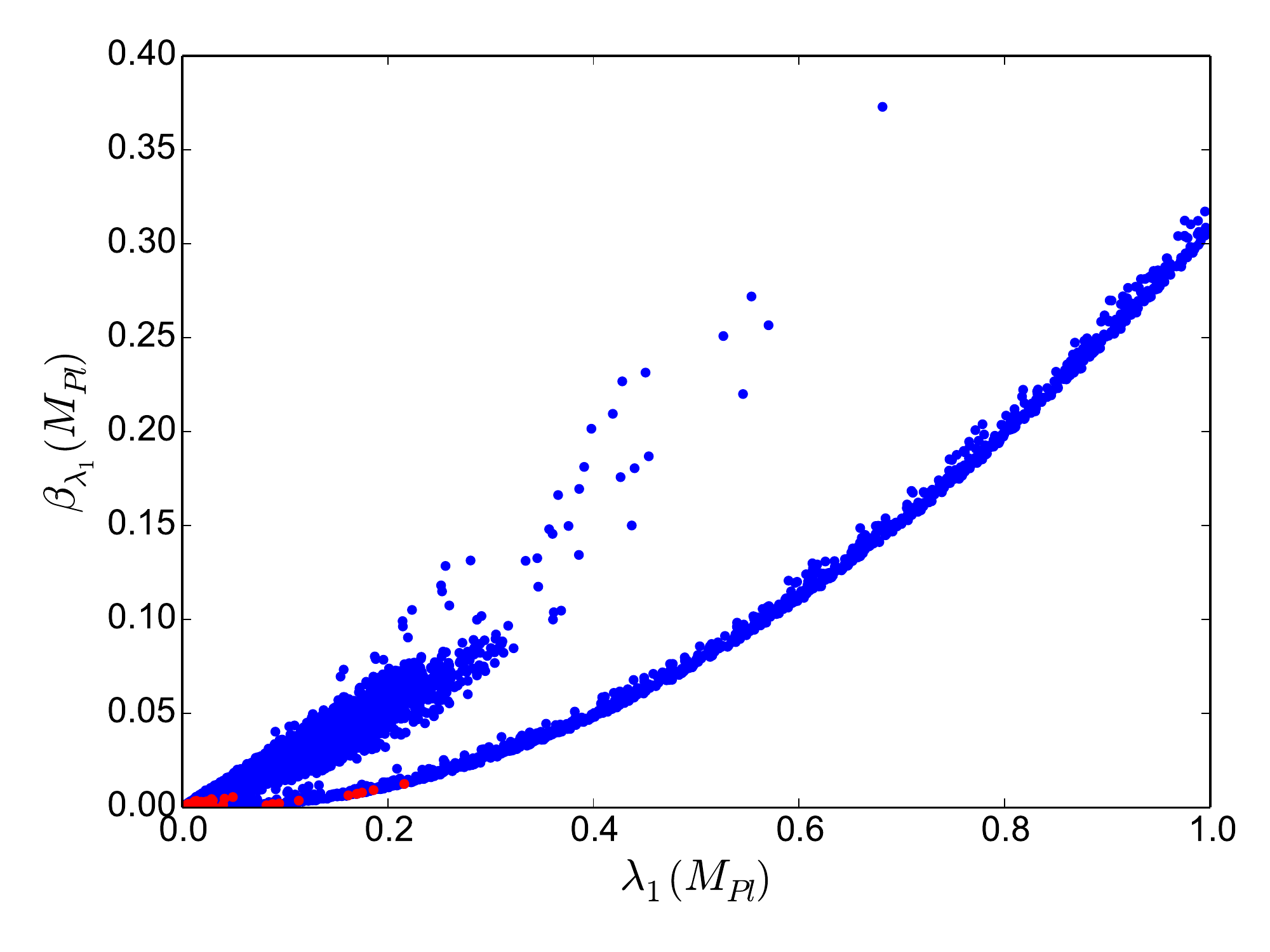}\label{fig:Inert_l1_betal1_comparison_theoretical}}
  \hfill
  \subfloat[]{\includegraphics[width=0.5\textwidth]{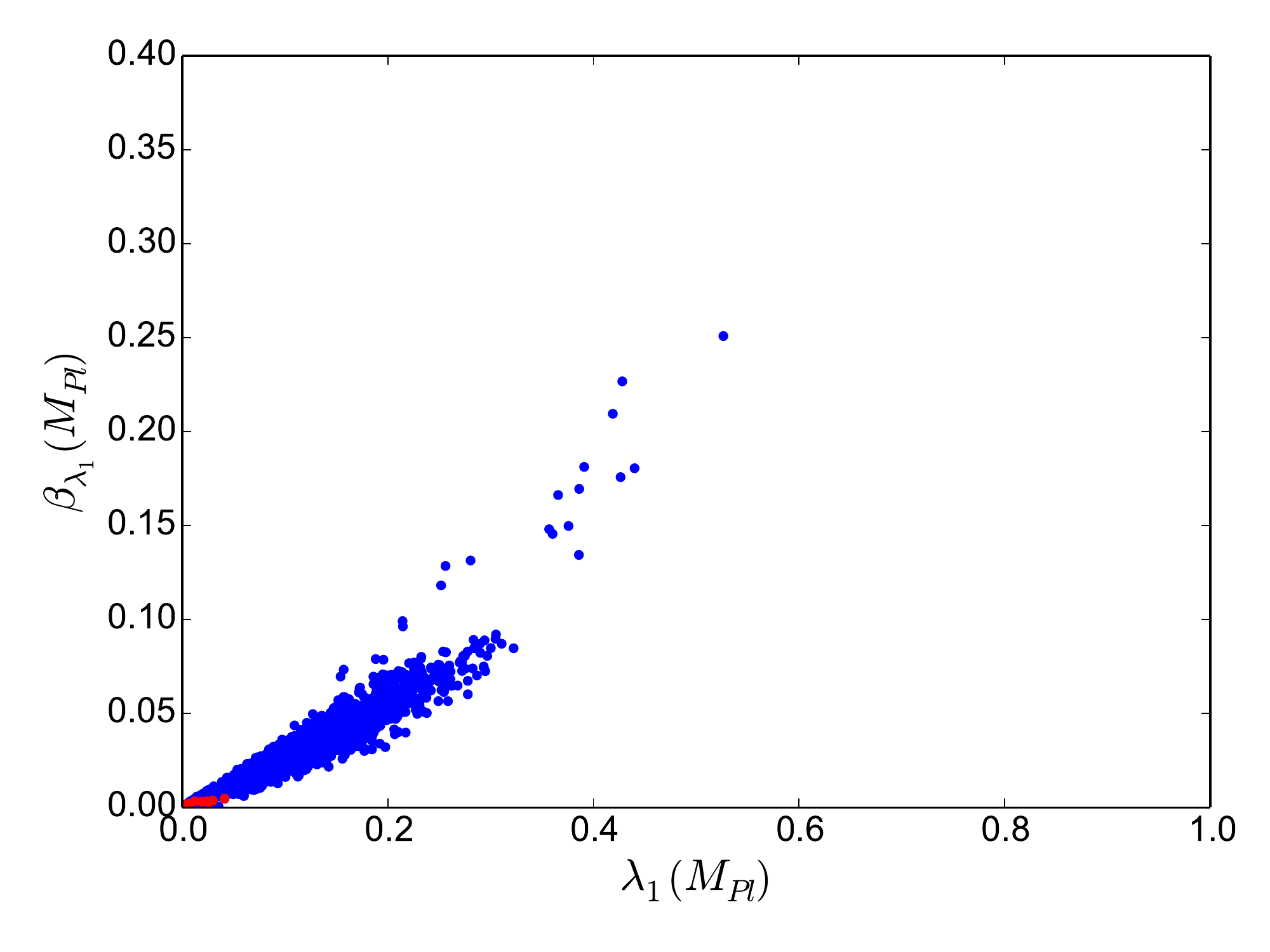}\label{fig:Inert_l1_betal1_comparison_experimental}}
\caption{Compatible values of the Higgs quartic coupling $\lambda_1 \left( M_{\rm Pl} \right)$ against $\beta_{\lambda_1} \left( M_{\rm Pl} \right)$ in the IDM. \textbf{(a)} includes points that are stable and perturbative up to $M_{Pl}$ and include an SM Higgs candidate, whilst \textbf{(b)} also enforces all relevant experimental constraints discussed in section \ref{sec:THDM_numerical_analysis_constraints}. Blue points obey $\beta_{\lambda_{1,2,3,4}} < 1.0$ at $M_{Pl}$ whilst red points obey $\beta_{\lambda_1} < 0.0127$, $\beta_{\lambda_2} < 0.0064$, $\beta_{\lambda_3} < 0.0139$, $\beta_{\lambda_4} < 0.0030$ at $M_{Pl}$. \\ \\}
\label{fig:Inert_l1_betal1_comparison}
\end{figure}

\begin{figure}[tbh]
  \centering
  \subfloat[]{\includegraphics[width=0.5\textwidth]{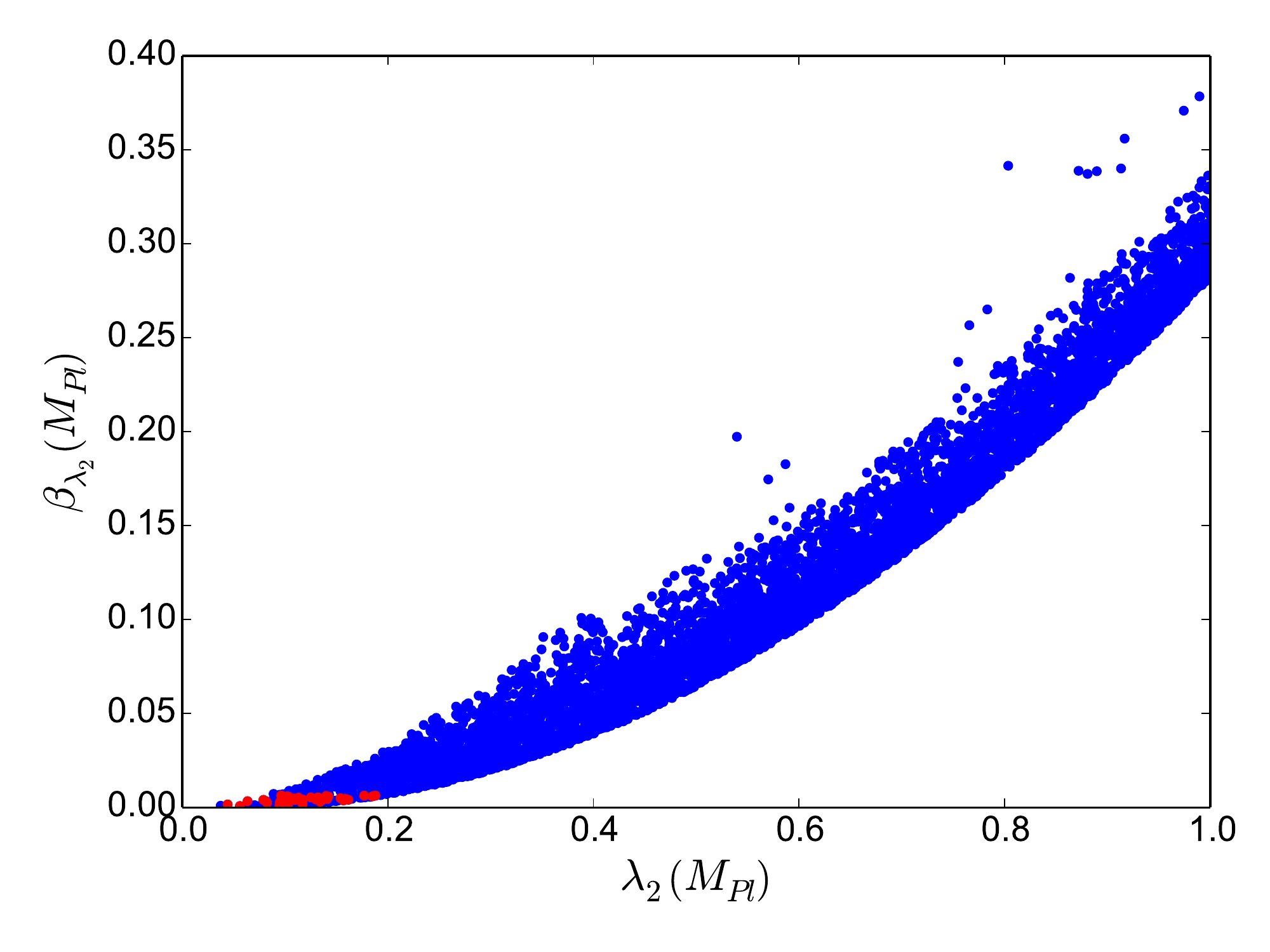}\label{fig:Inert_l2_betal2_comparison_theoretical}}
  \hfill
  \subfloat[]{\includegraphics[width=0.5\textwidth]{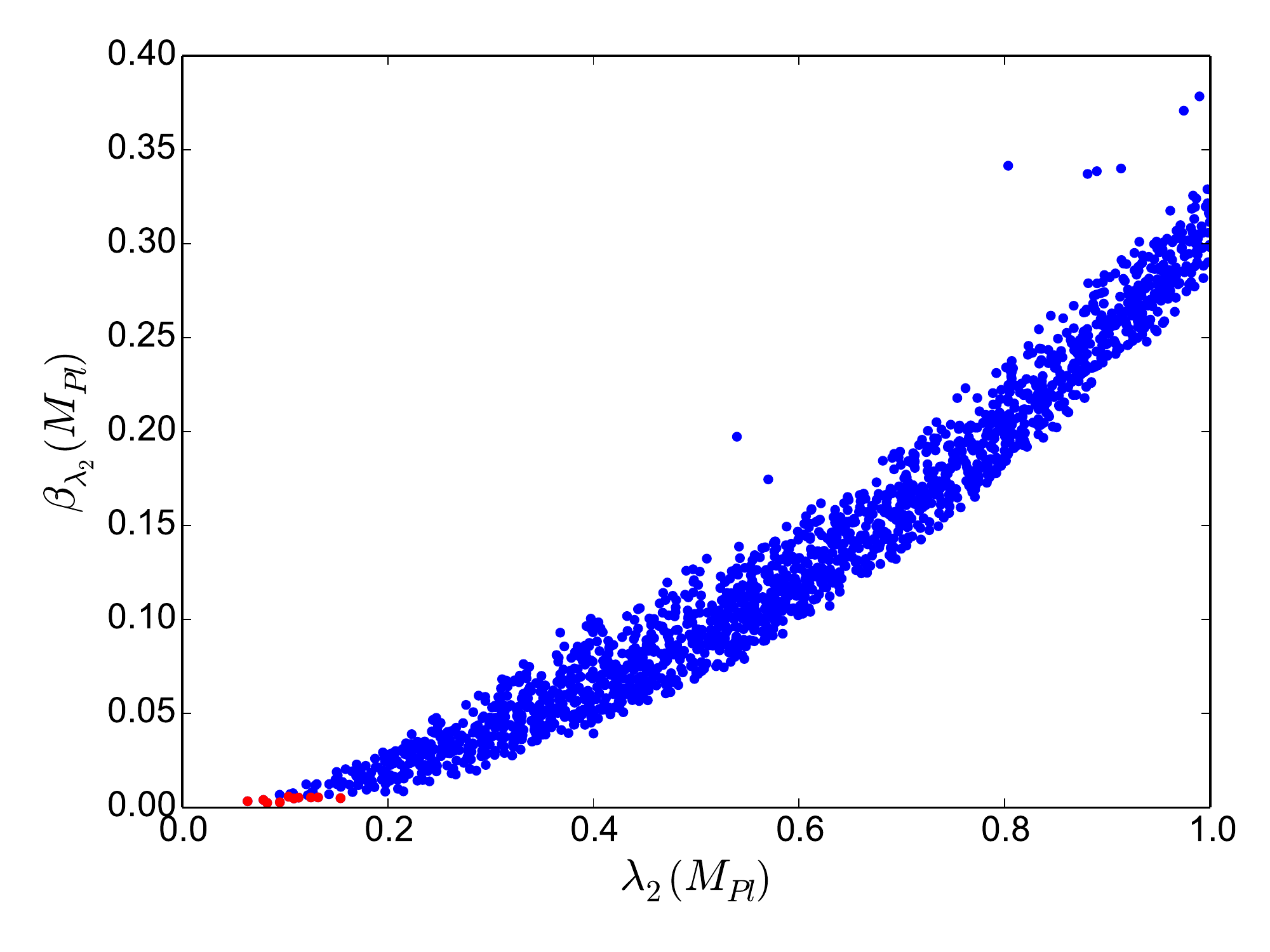}\label{fig:Inert_l2_betal2_comparison_experimental}}
\caption{Compatible values of the Higgs quartic coupling $\lambda_2 \left( M_{\rm Pl} \right)$ against $\beta_{\lambda_2} \left( M_{\rm Pl} \right)$ in the IDM. \textbf{(a)} includes points that are stable and perturbative up to $M_{Pl}$ and include an SM Higgs candidate, whilst \textbf{(b)} also enforces all relevant experimental constraints discussed in section \ref{sec:THDM_numerical_analysis_constraints}. Blue points obey $\beta_{\lambda_{1,2,3,4}} < 1.0$ at $M_{Pl}$ whilst red points obey $\beta_{\lambda_1} < 0.0127$, $\beta_{\lambda_2} < 0.0064$, $\beta_{\lambda_3} < 0.0139$, $\beta_{\lambda_4} < 0.0030$ at $M_{Pl}$. \\ \\}
\label{fig:Inert_l2_betal2_comparison}
\end{figure}

\begin{figure}[tbh]
  \centering
  \subfloat[]{\includegraphics[width=0.5\textwidth]{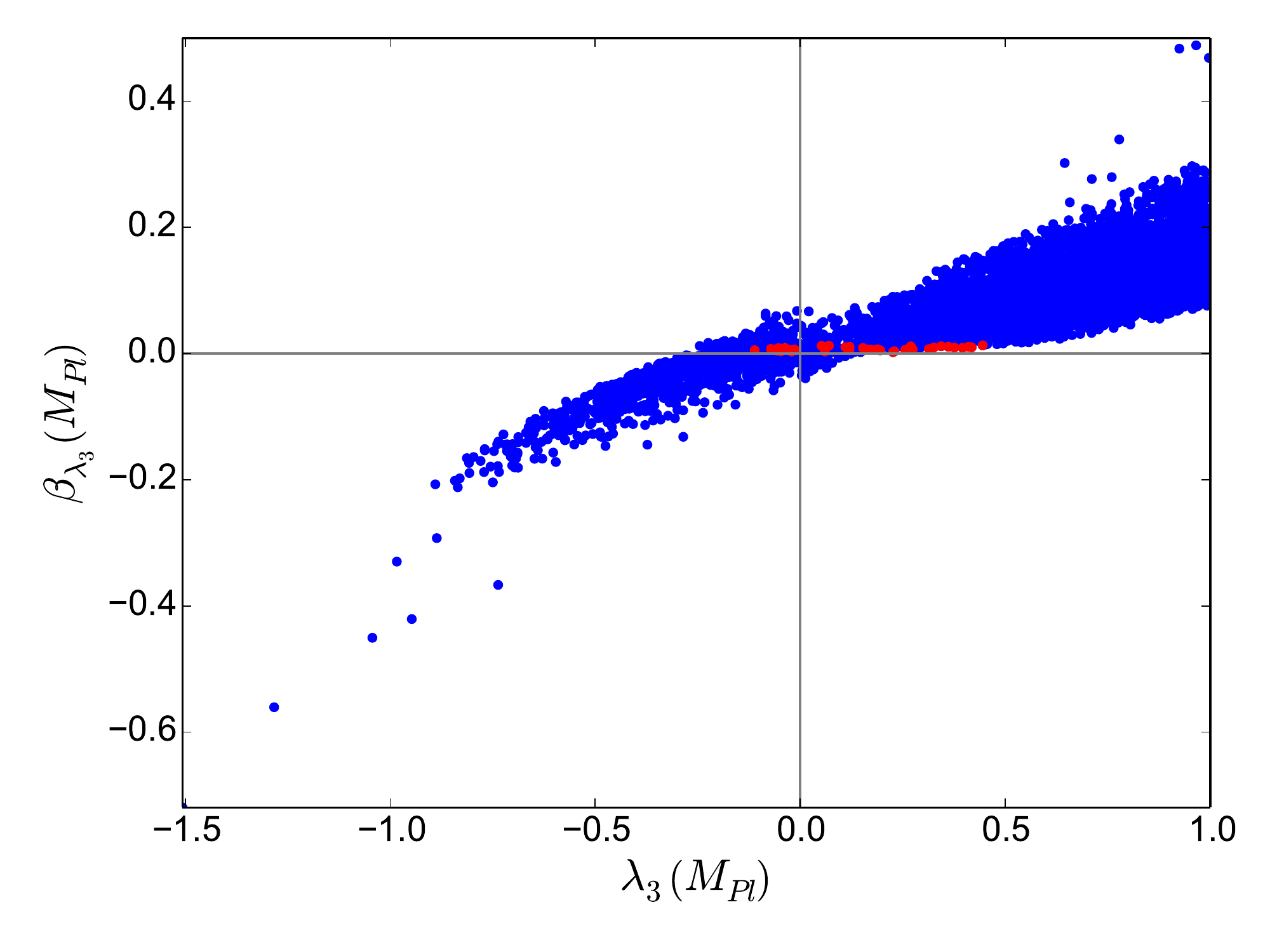}\label{fig:Inert_l3_betal3_comparison_theoretical}}
  \hfill
  \subfloat[]{\includegraphics[width=0.5\textwidth]{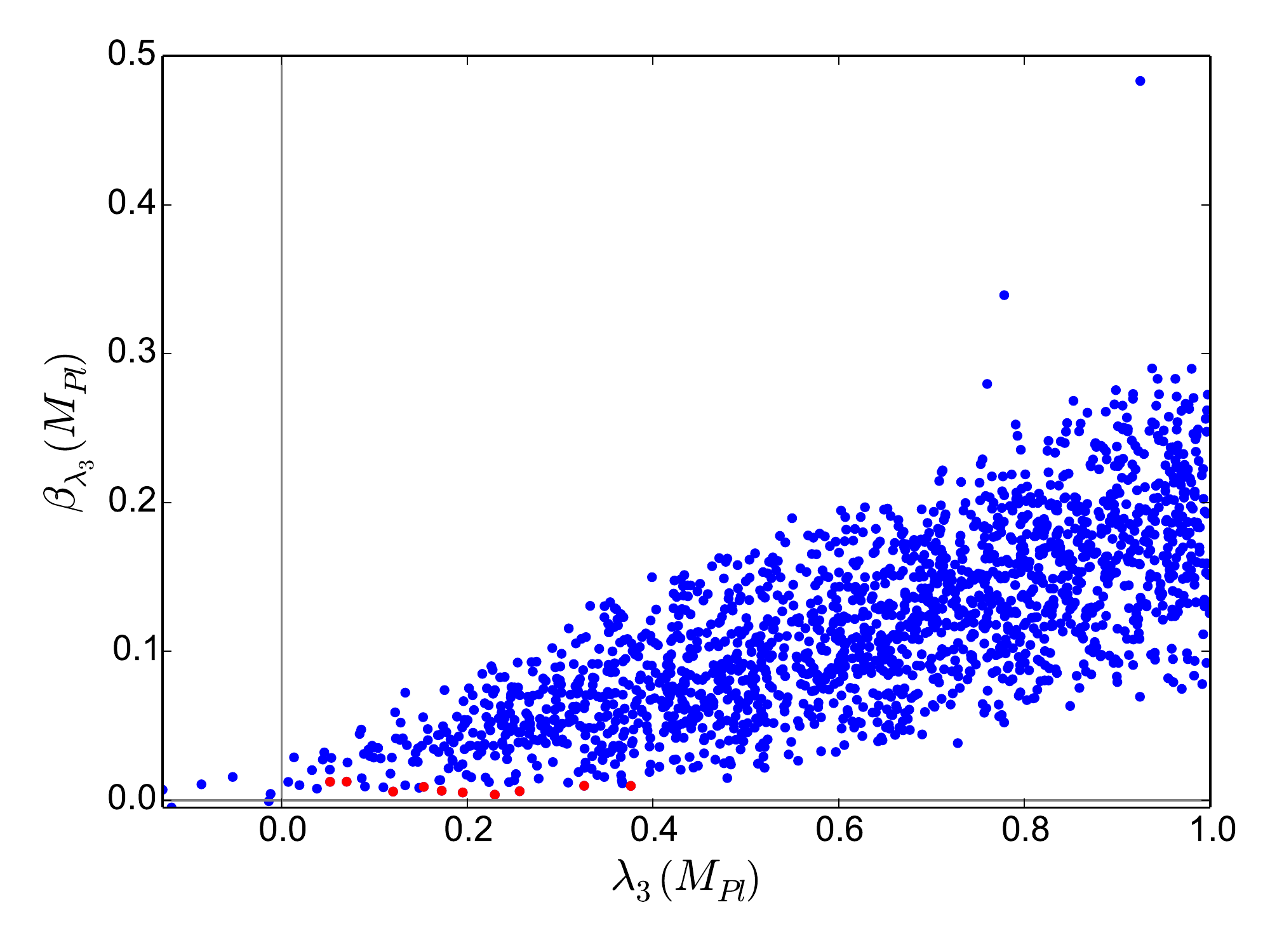}\label{fig:Inert_l3_betal3_comparison_experimental}}
\caption{Compatible values of the Higgs quartic coupling $\lambda_3 \left( M_{\rm Pl} \right)$ against $\beta_{\lambda_3} \left( M_{\rm Pl} \right)$ in the IDM. \textbf{(a)} includes points that are stable and perturbative up to $M_{Pl}$ and include an SM Higgs candidate, whilst \textbf{(b)} also enforces all relevant experimental constraints discussed in section \ref{sec:THDM_numerical_analysis_constraints}. Blue points obey $\beta_{\lambda_{1,2,3,4}} < 1.0$ at $M_{Pl}$ whilst red points obey $\beta_{\lambda_1} < 0.0127$, $\beta_{\lambda_2} < 0.0064$, $\beta_{\lambda_3} < 0.0139$, $\beta_{\lambda_4} < 0.0030$ at $M_{Pl}$. \\ \\}
\label{fig:Inert_l3_betal3_comparison}
\end{figure}

\begin{figure}[tbh]
  \centering
  \subfloat[]{\includegraphics[width=0.5\textwidth]{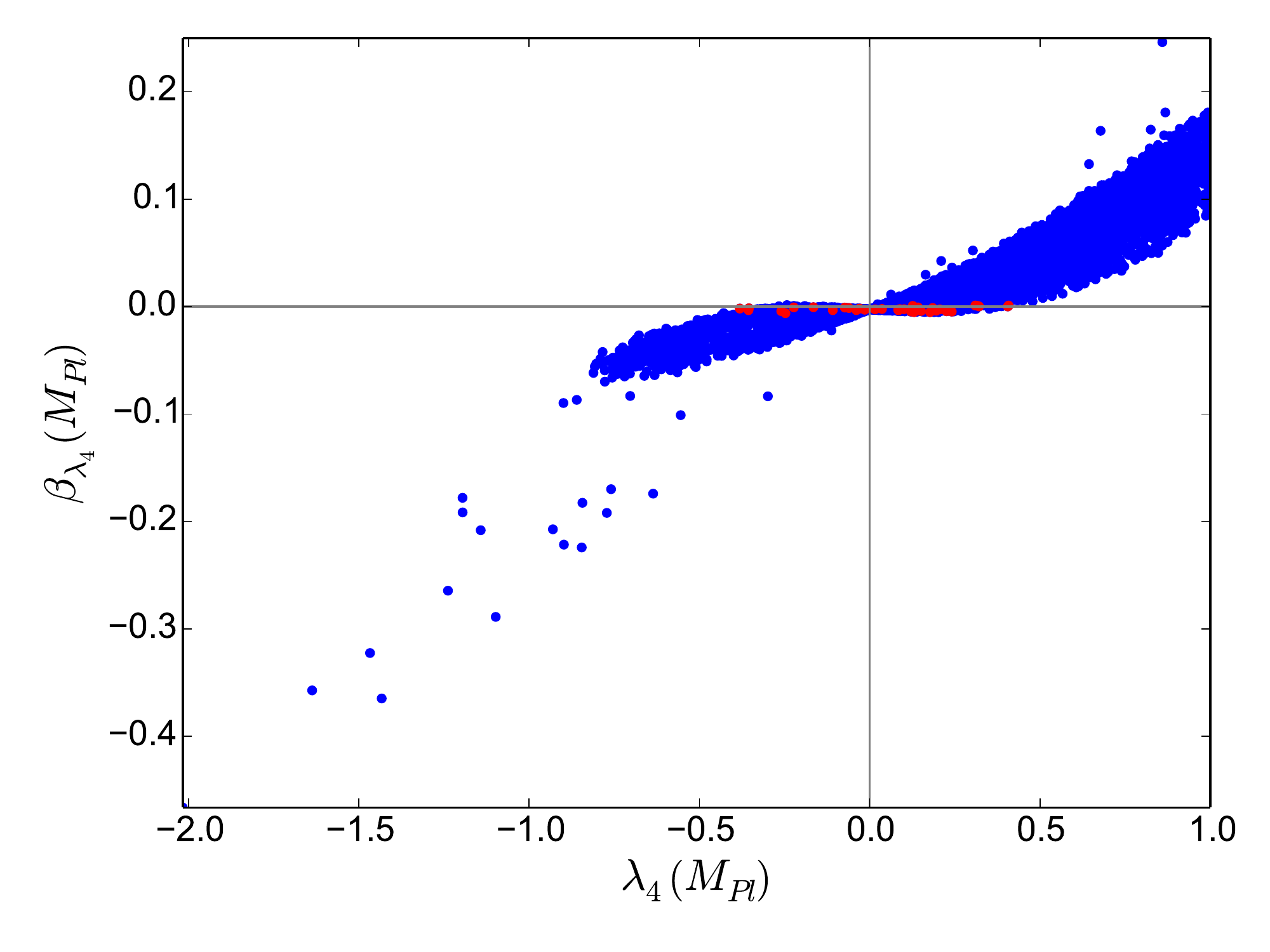}\label{fig:Inert_l4_betal4_comparison_theoretical}}
  \hfill
  \subfloat[]{\includegraphics[width=0.5\textwidth]{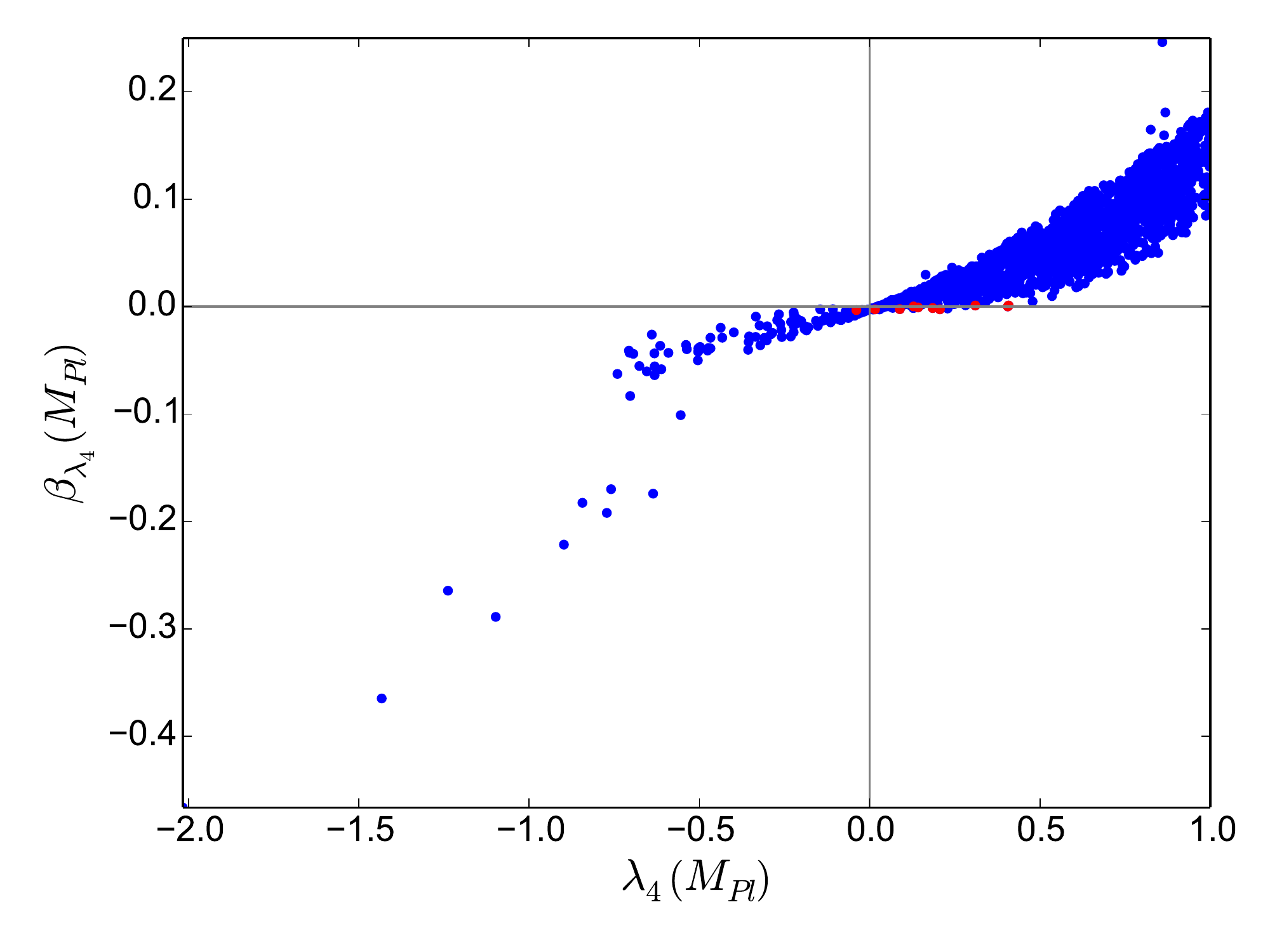}\label{fig:Inert_l4_betal4_comparison_experimental}}
\caption{Compatible values of the Higgs quartic coupling $\lambda_4 \left( M_{\rm Pl} \right)$ against $\beta_{\lambda_4} \left( M_{\rm Pl} \right)$ in the IDM. \textbf{(a)} includes points that are stable and perturbative up to $M_{Pl}$ and include an SM Higgs candidate, whilst \textbf{(b)} also enforces all relevant experimental constraints discussed in section \ref{sec:THDM_numerical_analysis_constraints}. Blue points obey $\beta_{\lambda_{1,2,3,4}} < 1.0$ at $M_{Pl}$ whilst red points obey $\beta_{\lambda_1} < 0.0127$, $\beta_{\lambda_2} < 0.0064$, $\beta_{\lambda_3} < 0.0139$, $\beta_{\lambda_4} < 0.0030$ at $M_{Pl}$. \\ \\}
\label{fig:Inert_l4_betal4_comparison}
\end{figure}

We now present the results of our numerical analysis of the IDM. \Cref{fig:Inert_l1_betal1_comparison,fig:Inert_l2_betal2_comparison,fig:Inert_l3_betal3_comparison,fig:Inert_l4_betal4_comparison} show points in the $\lambda_i - \beta_{\lambda_i}$ plane that satisfy both our theoretical and experimental constraints as well as the asymptotic safety high scale boundary conditions of Eq. \ref{eq:THDM_truncation_error}. The situation is somewhat similar to the Type-II case discussed in \ref{sec:THDM_AS}, inasmuch as there are points in the parameter space that are compatible with asymptotic safety  and that those points have very small values of the quartic couplings.

\begin{figure}[t!]
  \centering
  \subfloat[]{\includegraphics[width=0.5\textwidth]{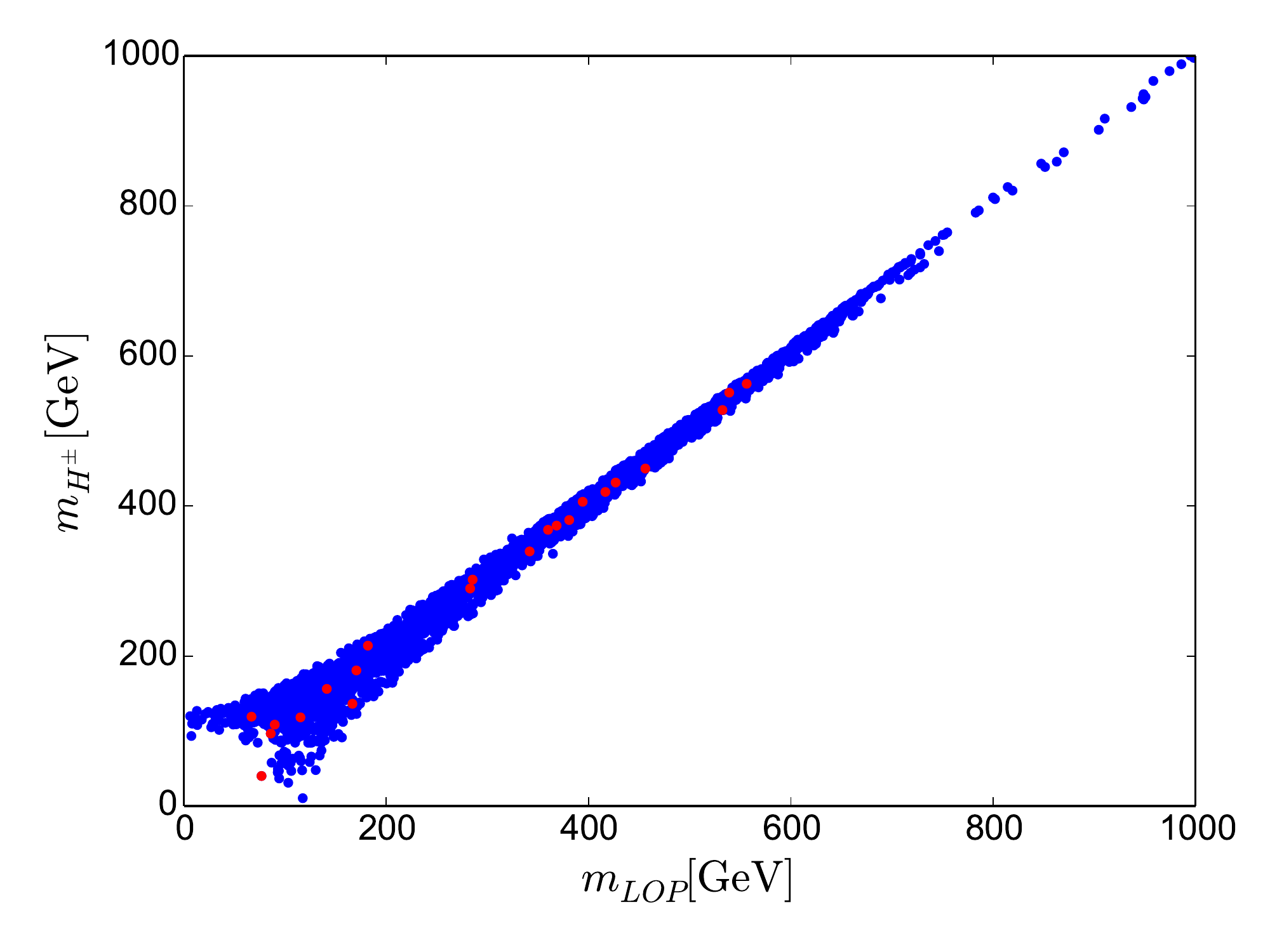}\label{fig:Inert_LOP_chargedhiggs_comparison_theoretical}}
  \hfill
  \subfloat[]{\includegraphics[width=0.5\textwidth]{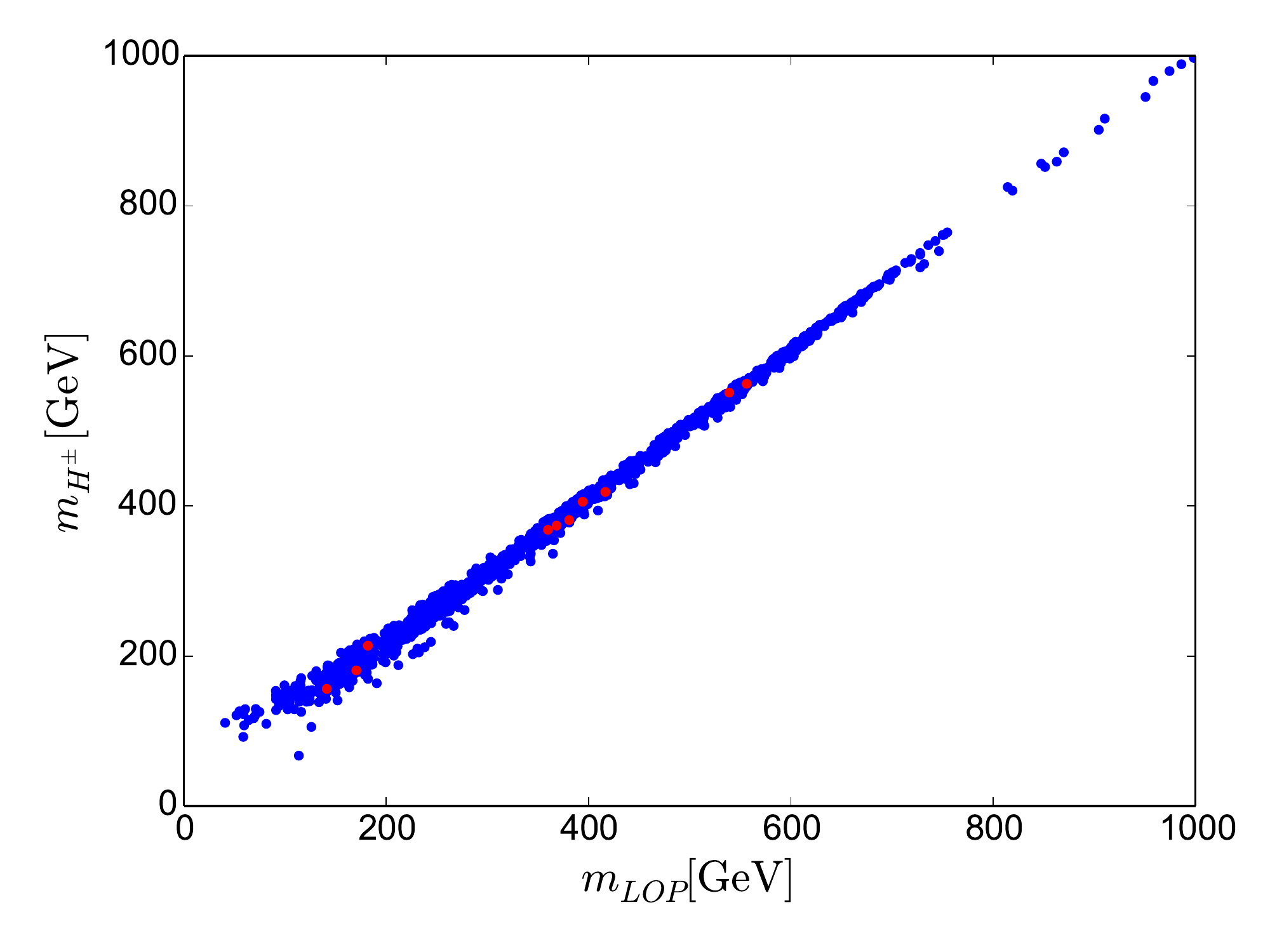}\label{fig:Inert_LOP_chargedhiggs_comparison_experimental}}
\caption{Compatible values of the Lightest Odd Particle mass $m_{LOP}$ against the charged Higgs mass $m_{H^{\pm}}$ in the IDM. \textbf{(a)} includes points that are stable and perturbative up to $M_{Pl}$ and include an SM Higgs candidate, whilst \textbf{(b)} also enforces all relevant experimental constraints discussed in section \ref{sec:THDM_numerical_analysis_constraints}. Blue points obey $\beta_{\lambda_{1,2,3,4}} < 1.0$ at $M_{Pl}$ whilst red points obey $\beta_{\lambda_1} < 0.0127$, $\beta_{\lambda_2} < 0.0064$, $\beta_{\lambda_3} < 0.0139$, $\beta_{\lambda_4} < 0.0030$ at $M_{Pl}$.}
\label{fig:Inert_LOP_chargedhiggs_comparison}
\end{figure}

Figure \ref{fig:Inert_LOP_chargedhiggs_comparison} shows the allowed masses of the DM candidate $m_{LOP}$ and the charged Higgs mass $m_{H^{\pm}}$. The requirement that the LOP account for the DM relic density and the results from DM direct detection experiments places a lower limit on the LOP mass of $m_{LOP} \approx 40\,$GeV. As for the Type-II case, points which meet the high scale constraint of asymptotic safety are seen to have a wide range of allowed scalar masses. It appears from our results that the existence of an interacting UV fixed point for the quartic couplings is valid under both the Type-II model and the Inert model. It places constraints on the high scale values of the quartic couplings, but due to the freedom to vary $m_{22}^2$ this does not translate to strong constraints on the possible masses of the new scalars.

\section{Conclusions}
\label{sec:conclusions}
We have investigated the Type-II 2HDM and the IDM with a focus on possible constraints on the quartic Higgs couplings and their $\beta$ functions at the Planck scale. These high scale conditions may be a consequence of a second minimum in the potential that is degenerate with the electroweak minimum, as is the case in the MPP, or they may be due to the couplings running towards an interacting UV fixed point at $M_{Pl}$, as for Asymptotic Safety. In this work we have examined the viability of these models with the required high-scale boundary conditions, checking  their compatibility with perturbativity, vacuum stability, and a SM Higgs candidate of the appropriate mass, as well as experimental constraints from colliders, flavour physics and DM experiments.

Models with a second Higgs doublet have much more flexibility in their scalar potential than models with only one Higgs doublet, which gives them more freedom to accommodate the boundary conditions of the the MPP or asymptotic safety. However, we found that both the Type-II 2HDM and the IDM cannot satisfy the conditions required at the Planck scale by the MPP. Specifically, we found no points in either model's parameter space that was consistent with the MPP whilst also having a valid SM Higgs, an experimentally acceptable top quark mass, and a stable vacuum. In the Type-II case we found that a stable vacuum would require a top mass on the order of $230\,$GeV, whilst in the Inert case we found no points at all that could meet our theoretical requirements. The results of our analysis would suggest that the MPP is not compatible with the 2HDM nor IDM that we investigated.

Asymptotic safety remains viable, as we found numerous points in the parameter space of both the 2HDM and IDM that were compatible with theoretical and experimental constraints and had the required Planck scale values of the quartic Higgs $\beta$-functions. These points also have small but non-zero values of the corresponding quartic couplings, which is entirely in keeping with the existence of an interacting UV fixed point. The type-II case has a lower limit on the masses of the additional scalars of $m_{H,A,H^{\pm}} \approx 330 \,$GeV imposed by experimental constraints. In the IDM the DM relic density and direct detection experiments place constraints on the mass of the model's DM candidate of $m_{LOP} \approx 40\,$GeV. Although our investigation found regions of parameter space that are compatible with all constraints, they correspond to a range of masses for the extra Higgs, with no apparent restriction on those masses coming from the high scale boundary conditions.

Of course, the non-viability of the MPP for these two models does not imply that is it wrong. One could imagine additional matter being added to the model that could make such scenarios viable again. Additional matter added to the 2HDM or IDM would have the difficult task of forcing $\tilde \lambda$ to stay positive. However, it would be interesting to examine the SM Higgs sector with alternative additions, such as vector-like fermions. Ultimately the question remains, is the peculiar behaviour of the SM Higgs potential at the Planck scale a coincidence or a sign of new physics?


\section*{Acknowledgements}

The authors would like to thank Peter Athron for invaluable help with FlexibleSUSY; as well as Karl Nordstrom, Ant\'onio Morais and David Sutherland for useful discussions. DJM acknowledges partial support from the STFC grants ST/L000446/1 and ST/P000746/1.

\begin{appendices}
\input{Appendix1}
\end{appendices}

\bibliography{bibliography}{}
\bibliographystyle{JHEP}

\end{document}

%% file: Appendix1.tex
\section{Renormalisation Group Equations of the Quartic Higgs Couplings}
\label{Appendix:RGEs}
The two-loop $\beta$-functions that describe the running of the quartic Higgs couplings $\lambda_{1-7}$ are calculated using SARAH \citep{Staub:2013tta}. We present here the one- and two-loop contributions to $\beta_{\lambda_i}$ ($i=1\ldots 7$), that is $\beta_{\lambda_i}^{(1)}$ and $\beta_{\lambda_i}^{(2)}$ respectively; $g_{1-3}$ are the SM gauge couplings and $Y_f$ ($f=\{u,d,e\}$) are the Yukawa matrices.
{\allowdisplaybreaks  \begin{align}
\beta_{\lambda_1}^{(1)} & =
+\frac{27}{200} g_{1}^{4} +\frac{9}{20} g_{1}^{2} g_{2}^{2} +\frac{9}{8} g_{2}^{4} -\frac{9}{5} g_{1}^{2} \lambda_1 -9 g_{2}^{2} \lambda_1 +24 \lambda_{1}^{2} +2 \lambda_{3}^{2} +2 \lambda_3 \lambda_4 +\lambda_{4}^{2}+\lambda_{5}^{2}+12 \lambda_{6}^{2} \nonumber \\
 &+12 \lambda_1 \mbox{Tr}\Big({Y_d  Y_{d}^{\dagger}}\Big) +4 \lambda_1 \mbox{Tr}\Big({Y_e  Y_{e}^{\dagger}}\Big) -6 \mbox{Tr}\Big({Y_d  Y_{d}^{\dagger}  Y_d  Y_{d}^{\dagger}}\Big) -2 \mbox{Tr}\Big({Y_e  Y_{e}^{\dagger}  Y_e  Y_{e}^{\dagger}}\Big) \\
\beta_{\lambda_1}^{(2)} & =
-\frac{3537}{2000} g_{1}^{6} -\frac{1719}{400} g_{1}^{4} g_{2}^{2} -\frac{303}{80} g_{1}^{2} g_{2}^{4} +\frac{291}{16} g_{2}^{6} +\frac{1953}{200} g_{1}^{4} \lambda_1 +\frac{117}{20} g_{1}^{2} g_{2}^{2} \lambda_1 -\frac{51}{8} g_{2}^{4} \lambda_1 \\ \nonumber &+\frac{108}{5} g_{1}^{2} \lambda_{1}^{2} +108 g_{2}^{2} \lambda_{1}^{2} -312 \lambda_{1}^{3} +\frac{9}{10} g_{1}^{4} \lambda_3 +\frac{15}{2} g_{2}^{4} \lambda_3 +\frac{12}{5} g_{1}^{2} \lambda_{3}^{2} +12 g_{2}^{2} \lambda_{3}^{2} -20 \lambda_1 \lambda_{3}^{2} -8 \lambda_{3}^{3} \nonumber \\ 
&+\frac{9}{20} g_{1}^{4} \lambda_4 +\frac{3}{2} g_{1}^{2} g_{2}^{2} \lambda_4 +\frac{15}{4} g_{2}^{4} \lambda_4 +\frac{12}{5} g_{1}^{2} \lambda_3 \lambda_4 +12 g_{2}^{2} \lambda_3 \lambda_4 -20 \lambda_1 \lambda_3 \lambda_4 -12 \lambda_{3}^{2} \lambda_4 +\frac{6}{5} g_{1}^{2} \lambda_{4}^{2} \nonumber \\
 &+3 g_{2}^{2} \lambda_{4}^{2} -12 \lambda_1 \lambda_{4}^{2} -16 \lambda_3 \lambda_{4}^{2} -6 \lambda_{4}^{3} -\frac{3}{5} g_{1}^{2} \lambda_{5}^{2} -14 \lambda_1 \lambda_{5}^{2} -20 \lambda_3 \lambda_{5}^{2} -22 \lambda_4 \lambda_{5}^{2} \nonumber \\
 &+\frac{54}{5} g_{1}^{2} \lambda_{6}^{2} +54 g_{2}^{2} \lambda_{6}^{2} -318 \lambda_1 \lambda_{6}^{2} -66 \lambda_3 \lambda_{6}^{2} -70 \lambda_4 \lambda_{6}^{2} -74 \lambda_5 \lambda_{6}^{2} -36 \lambda_3 \lambda_6 \lambda_7 \nonumber \\ &-28 \lambda_4 \lambda_6 \lambda_7 -20 \lambda_5 \lambda_6 \lambda_7 +6 \lambda_1 \lambda_{7}^{2} -18 \lambda_3 \lambda_{7}^{2} -14 \lambda_4 \lambda_{7}^{2} -10 \lambda_5 \lambda_{7}^{2} \nonumber \\
 &+\frac{1}{20} \Big(-5 \Big(144 \lambda_{6}^{2}  -320 g_{3}^{2} \lambda_1  + 576 \lambda_{1}^{2}  -90 g_{2}^{2} \lambda_1  + 9 g_{2}^{4} \Big) + 9 g_{1}^{4}  + g_{1}^{2} \Big(50 \lambda_1  + 54 g_{2}^{2} \Big)\Big)\mbox{Tr}\Big({Y_d  Y_{d}^{\dagger}}\Big) \nonumber \\
 &-\frac{3}{20} \Big(15 g_{1}^{4}  -2 g_{1}^{2} \Big(11 g_{2}^{2}  + 25 \lambda_1 \Big) + 5 \Big(-10 g_{2}^{2} \lambda_1  + 16 \lambda_{6}^{2}  + 64 \lambda_{1}^{2}  + g_{2}^{4}\Big)\Big)\mbox{Tr}\Big({Y_e  Y_{e}^{\dagger}}\Big) \nonumber \\ 
 &-12 \lambda_{3}^{2} \mbox{Tr}\Big({Y_u  Y_{u}^{\dagger}}\Big) -12 \lambda_3 \lambda_4 \mbox{Tr}\Big({Y_u  Y_{u}^{\dagger}}\Big) -6 \lambda_{4}^{2} \mbox{Tr}\Big({Y_u  Y_{u}^{\dagger}}\Big) -6 \lambda_{5}^{2} \mbox{Tr}\Big({Y_u  Y_{u}^{\dagger}}\Big) -36 \lambda_{6}^{2} \mbox{Tr}\Big({Y_u  Y_{u}^{\dagger}}\Big) \nonumber \\
 &+\frac{4}{5} g_{1}^{2} \mbox{Tr}\Big({Y_d  Y_{d}^{\dagger}  Y_d  Y_{d}^{\dagger}}\Big) -32 g_{3}^{2} \mbox{Tr}\Big({Y_d  Y_{d}^{\dagger}  Y_d  Y_{d}^{\dagger}}\Big) -3 \lambda_1 \mbox{Tr}\Big({Y_d  Y_{d}^{\dagger}  Y_d  Y_{d}^{\dagger}}\Big) -9 \lambda_1 \mbox{Tr}\Big({Y_d  Y_{u}^{\dagger}  Y_u  Y_{d}^{\dagger}}\Big) \nonumber \\
 &-\frac{12}{5} g_{1}^{2} \mbox{Tr}\Big({Y_e  Y_{e}^{\dagger}  Y_e  Y_{e}^{\dagger}}\Big) - \lambda_1 \mbox{Tr}\Big({Y_e  Y_{e}^{\dagger}  Y_e  Y_{e}^{\dagger}}\Big) +30 \mbox{Tr}\Big({Y_d  Y_{d}^{\dagger}  Y_d  Y_{d}^{\dagger}  Y_d  Y_{d}^{\dagger}}\Big) +6 \mbox{Tr}\Big({Y_d  Y_{u}^{\dagger}  Y_u  Y_{d}^{\dagger}  Y_d  Y_{d}^{\dagger}}\Big) \nonumber \\
 \beta_{\lambda_2}^{(1)} & =
 +\frac{27}{200} g_{1}^{4} +\frac{9}{20} g_{1}^{2} g_{2}^{2} +\frac{9}{8} g_{2}^{4} -\frac{9}{5} g_{1}^{2} \lambda_2 -9 g_{2}^{2} \lambda_2 +24 \lambda_{2}^{2} +2 \lambda_{3}^{2} +2 \lambda_3 \lambda_4 +\lambda_{4}^{2}+\lambda_{5}^{2}+12 \lambda_{7}^{2} \nonumber \\
  &+12 \lambda_2 \mbox{Tr}\Big({Y_u  Y_{u}^{\dagger}}\Big) -6 \mbox{Tr}\Big({Y_u  Y_{u}^{\dagger}  Y_u  Y_{u}^{\dagger}}\Big) \\
 \beta_{\lambda_2}^{(2)} & =
 -\frac{3537}{2000} g_{1}^{6} -\frac{1719}{400} g_{1}^{4} g_{2}^{2} -\frac{303}{80} g_{1}^{2} g_{2}^{4} +\frac{291}{16} g_{2}^{6} +\frac{1953}{200} g_{1}^{4} \lambda_2 +\frac{117}{20} g_{1}^{2} g_{2}^{2} \lambda_2 -\frac{51}{8} g_{2}^{4} \lambda_2 +\frac{108}{5} g_{1}^{2} \lambda_{2}^{2} \nonumber \\
  &+108 g_{2}^{2} \lambda_{2}^{2} -312 \lambda_{2}^{3} +\frac{9}{10} g_{1}^{4} \lambda_3 +\frac{15}{2} g_{2}^{4} \lambda_3 +\frac{12}{5} g_{1}^{2} \lambda_{3}^{2} +12 g_{2}^{2} \lambda_{3}^{2} -20 \lambda_2 \lambda_{3}^{2} -8 \lambda_{3}^{3} +\frac{9}{20} g_{1}^{4} \lambda_4 \nonumber \\
  &+\frac{3}{2} g_{1}^{2} g_{2}^{2} \lambda_4 +\frac{15}{4} g_{2}^{4} \lambda_4 +\frac{12}{5} g_{1}^{2} \lambda_3 \lambda_4 +12 g_{2}^{2} \lambda_3 \lambda_4 -20 \lambda_2 \lambda_3 \lambda_4 -12 \lambda_{3}^{2} \lambda_4 +\frac{6}{5} g_{1}^{2} \lambda_{4}^{2} \nonumber \\
  &+3 g_{2}^{2} \lambda_{4}^{2} -12 \lambda_2 \lambda_{4}^{2} -16 \lambda_3 \lambda_{4}^{2} -6 \lambda_{4}^{3} -\frac{3}{5} g_{1}^{2} \lambda_{5}^{2} -14 \lambda_2 \lambda_{5}^{2} -20 \lambda_3 \lambda_{5}^{2} -22 \lambda_4 \lambda_{5}^{2} +6 \lambda_2 \lambda_{6}^{2} \nonumber \\
  &-18 \lambda_3 \lambda_{6}^{2} -14 \lambda_4 \lambda_{6}^{2} -10 \lambda_5 \lambda_{6}^{2} -36 \lambda_3 \lambda_6 \lambda_7 -28 \lambda_4 \lambda_6 \lambda_7 -20 \lambda_5 \lambda_6 \lambda_7 +\frac{54}{5} g_{1}^{2} \lambda_{7}^{2} +54 g_{2}^{2} \lambda_{7}^{2} \nonumber \\
  &-318 \lambda_2 \lambda_{7}^{2} -66 \lambda_3 \lambda_{7}^{2} -70 \lambda_4 \lambda_{7}^{2} -74 \lambda_5 \lambda_{7}^{2} -6 \Big(2 \lambda_{3}^{2}  + 2 \lambda_3 \lambda_4  + 6 \lambda_{7}^{2}  + \lambda_{4}^{2} + \lambda_{5}^{2}\Big)\mbox{Tr}\Big({Y_d  Y_{d}^{\dagger}}\Big) \nonumber \\
  &-2 \Big(2 \lambda_{3}^{2}  + 2 \lambda_3 \lambda_4  + 6 \lambda_{7}^{2}  + \lambda_{4}^{2} + \lambda_{5}^{2}\Big)\mbox{Tr}\Big({Y_e  Y_{e}^{\dagger}}\Big) -\frac{171}{100} g_{1}^{4} \mbox{Tr}\Big({Y_u  Y_{u}^{\dagger}}\Big) +\frac{63}{10} g_{1}^{2} g_{2}^{2} \mbox{Tr}\Big({Y_u  Y_{u}^{\dagger}}\Big) \nonumber \\
  &-\frac{9}{4} g_{2}^{4} \mbox{Tr}\Big({Y_u  Y_{u}^{\dagger}}\Big) +\frac{17}{2} g_{1}^{2} \lambda_2 \mbox{Tr}\Big({Y_u  Y_{u}^{\dagger}}\Big) +\frac{45}{2} g_{2}^{2} \lambda_2 \mbox{Tr}\Big({Y_u  Y_{u}^{\dagger}}\Big) +80 g_{3}^{2} \lambda_2 \mbox{Tr}\Big({Y_u  Y_{u}^{\dagger}}\Big) \nonumber \\
  &-144 \lambda_{2}^{2} \mbox{Tr}\Big({Y_u  Y_{u}^{\dagger}}\Big) -36 \lambda_{7}^{2} \mbox{Tr}\Big({Y_u  Y_{u}^{\dagger}}\Big) -9 \lambda_2 \mbox{Tr}\Big({Y_d  Y_{u}^{\dagger}  Y_u  Y_{d}^{\dagger}}\Big) -\frac{8}{5} g_{1}^{2} \mbox{Tr}\Big({Y_u  Y_{u}^{\dagger}  Y_u  Y_{u}^{\dagger}}\Big) \nonumber \\
  &-32 g_{3}^{2} \mbox{Tr}\Big({Y_u  Y_{u}^{\dagger}  Y_u  Y_{u}^{\dagger}}\Big) -3 \lambda_2 \mbox{Tr}\Big({Y_u  Y_{u}^{\dagger}  Y_u  Y_{u}^{\dagger}}\Big) +6 \mbox{Tr}\Big({Y_d  Y_{u}^{\dagger}  Y_u  Y_{u}^{\dagger}  Y_u  Y_{d}^{\dagger}}\Big) \\ \nonumber 
  &+30 \mbox{Tr}\Big({Y_u  Y_{u}^{\dagger}  Y_u  Y_{u}^{\dagger}  Y_u  Y_{u}^{\dagger}}\Big)+10 \mbox{Tr}\Big({Y_e  Y_{e}^{\dagger}  Y_e  Y_{e}^{\dagger}  Y_e  Y_{e}^{\dagger}}\Big) \\
 \beta_{\lambda_3}^{(1)} & =
 +\frac{27}{100} g_{1}^{4} -\frac{9}{10} g_{1}^{2} g_{2}^{2} +\frac{9}{4} g_{2}^{4} -\frac{9}{5} g_{1}^{2} \lambda_3 -9 g_{2}^{2} \lambda_3 +12 \lambda_1 \lambda_3 +12 \lambda_2 \lambda_3 +4 \lambda_{3}^{2} +4 \lambda_1 \lambda_4 \nonumber \\ 
 &+4 \lambda_2 \lambda_4 +2 \lambda_{4}^{2} +2 \lambda_{5}^{2} +4 \lambda_{6}^{2} +16 \lambda_6 \lambda_7 +4 \lambda_{7}^{2} +6 \lambda_3 \mbox{Tr}\Big({Y_d  Y_{d}^{\dagger}}\Big) +2 \lambda_3 \mbox{Tr}\Big({Y_e  Y_{e}^{\dagger}}\Big) \nonumber \\ 
 &+6 \lambda_3 \mbox{Tr}\Big({Y_u  Y_{u}^{\dagger}}\Big) -12 \mbox{Tr}\Big({Y_d  Y_{u}^{\dagger}  Y_u  Y_{d}^{\dagger}}\Big) \\
 \beta_{\lambda_3}^{(2)} & =
 -\frac{3537}{1000} g_{1}^{6} +\frac{909}{200} g_{1}^{4} g_{2}^{2} +\frac{33}{40} g_{1}^{2} g_{2}^{4} +\frac{291}{8} g_{2}^{6} +\frac{27}{10} g_{1}^{4} \lambda_1 -3 g_{1}^{2} g_{2}^{2} \lambda_1 +\frac{45}{2} g_{2}^{4} \lambda_1 +\frac{27}{10} g_{1}^{4} \lambda_2 \nonumber \\
  &-3 g_{1}^{2} g_{2}^{2} \lambda_2 +\frac{45}{2} g_{2}^{4} \lambda_2 +\frac{1773}{200} g_{1}^{4} \lambda_3 +\frac{33}{20} g_{1}^{2} g_{2}^{2} \lambda_3 -\frac{111}{8} g_{2}^{4} \lambda_3 +\frac{72}{5} g_{1}^{2} \lambda_1 \lambda_3 +72 g_{2}^{2} \lambda_1 \lambda_3 \nonumber \\
  &-60 \lambda_{1}^{2} \lambda_3 +\frac{72}{5} g_{1}^{2} \lambda_2 \lambda_3 +72 g_{2}^{2} \lambda_2 \lambda_3 -60 \lambda_{2}^{2} \lambda_3 +\frac{6}{5} g_{1}^{2} \lambda_{3}^{2} +6 g_{2}^{2} \lambda_{3}^{2} -72 \lambda_1 \lambda_{3}^{2} -72 \lambda_2 \lambda_{3}^{2} \nonumber \\
  &-12 \lambda_{3}^{3} +\frac{9}{10} g_{1}^{4} \lambda_4 -\frac{9}{5} g_{1}^{2} g_{2}^{2} \lambda_4 +\frac{15}{2} g_{2}^{4} \lambda_4 +\frac{24}{5} g_{1}^{2} \lambda_1 \lambda_4 +36 g_{2}^{2} \lambda_1 \lambda_4 -16 \lambda_{1}^{2} \lambda_4 +\frac{24}{5} g_{1}^{2} \lambda_2 \lambda_4 \nonumber \\
  &+36 g_{2}^{2} \lambda_2 \lambda_4 -16 \lambda_{2}^{2} \lambda_4 -12 g_{2}^{2} \lambda_3 \lambda_4 -32 \lambda_1 \lambda_3 \lambda_4 -32 \lambda_2 \lambda_3 \lambda_4 -4 \lambda_{3}^{2} \lambda_4 -\frac{6}{5} g_{1}^{2} \lambda_{4}^{2} \nonumber \\
  &+6 g_{2}^{2} \lambda_{4}^{2} -28 \lambda_1 \lambda_{4}^{2} -28 \lambda_2 \lambda_{4}^{2} -16 \lambda_3 \lambda_{4}^{2} -12 \lambda_{4}^{3} +\frac{12}{5} g_{1}^{2} \lambda_{5}^{2} -36 \lambda_1 \lambda_{5}^{2} -36 \lambda_2 \lambda_{5}^{2} \nonumber \\
  &-18 \lambda_3 \lambda_{5}^{2} -44 \lambda_4 \lambda_{5}^{2} +\frac{6}{5} g_{1}^{2} \lambda_{6}^{2} -124 \lambda_1 \lambda_{6}^{2} -44 \lambda_2 \lambda_{6}^{2} -60 \lambda_3 \lambda_{6}^{2} -68 \lambda_4 \lambda_{6}^{2} -68 \lambda_5 \lambda_{6}^{2} \nonumber \\
  &+\frac{96}{5} g_{1}^{2} \lambda_6 \lambda_7 +108 g_{2}^{2} \lambda_6 \lambda_7 -88 \lambda_1 \lambda_6 \lambda_7 -88 \lambda_2 \lambda_6 \lambda_7 -176 \lambda_3 \lambda_6 \lambda_7 -88 \lambda_4 \lambda_6 \lambda_7 \nonumber \\ 
 &-72 \lambda_5 \lambda_6 \lambda_7+\frac{6}{5} g_{1}^{2} \lambda_{7}^{2} -44 \lambda_1 \lambda_{7}^{2} -124 \lambda_2 \lambda_{7}^{2} -60 \lambda_3 \lambda_{7}^{2} -68 \lambda_4 \lambda_{7}^{2} -68 \lambda_5 \lambda_{7}^{2} \nonumber \\
  &+\frac{1}{20} \Big(9 g_{1}^{4} +g_{1}^{2} \Big(25 \lambda_3  -54 g_{2}^{2} \Big) -5 \Big(-45 g_{2}^{2} \lambda_3 \nonumber \\  
  &+ 8 \Big(-20 g_{3}^{2} \lambda_3  + 3 \Big(2 \lambda_{3}^{2}  + 4 \lambda_1 \Big(3 \lambda_3  + \lambda_4\Big) + 4 \lambda_{6}^{2}  + 8 \lambda_6 \lambda_7  + \lambda_{4}^{2} + \lambda_{5}^{2}\Big)\Big) + 9 g_{2}^{4} \Big)\Big)\mbox{Tr}\Big({Y_d  Y_{d}^{\dagger}}\Big) \nonumber \\
  &-\frac{1}{20} \Big(45 g_{1}^{4}  + 5 \Big(-15 g_{2}^{2} \lambda_3  + 3 g_{2}^{4}  + 8 \Big(2 \lambda_{3}^{2}  + 4 \lambda_1 \Big(3 \lambda_3  + \lambda_4\Big) + 4 \lambda_{6}^{2}  + 8 \lambda_6 \lambda_7  + \lambda_{4}^{2} + \lambda_{5}^{2}\Big)\Big) \nonumber \\
  &+ g_{1}^{2} \Big(66 g_{2}^{2}  -75 \lambda_3 \Big)\Big)\mbox{Tr}\Big({Y_e  Y_{e}^{\dagger}}\Big) -\frac{171}{100} g_{1}^{4} \mbox{Tr}\Big({Y_u  Y_{u}^{\dagger}}\Big) -\frac{63}{10} g_{1}^{2} g_{2}^{2} \mbox{Tr}\Big({Y_u  Y_{u}^{\dagger}}\Big) -\frac{9}{4} g_{2}^{4} \mbox{Tr}\Big({Y_u  Y_{u}^{\dagger}}\Big) \nonumber \\
  &+\frac{17}{4} g_{1}^{2} \lambda_3 \mbox{Tr}\Big({Y_u  Y_{u}^{\dagger}}\Big) +\frac{45}{4} g_{2}^{2} \lambda_3 \mbox{Tr}\Big({Y_u  Y_{u}^{\dagger}}\Big) +40 g_{3}^{2} \lambda_3 \mbox{Tr}\Big({Y_u  Y_{u}^{\dagger}}\Big) -72 \lambda_2 \lambda_3 \mbox{Tr}\Big({Y_u  Y_{u}^{\dagger}}\Big) \nonumber \\ 
  &-12 \lambda_{3}^{2} \mbox{Tr}\Big({Y_u  Y_{u}^{\dagger}}\Big) -24 \lambda_2 \lambda_4 \mbox{Tr}\Big({Y_u  Y_{u}^{\dagger}}\Big) -6 \lambda_{4}^{2} \mbox{Tr}\Big({Y_u  Y_{u}^{\dagger}}\Big) -6 \lambda_{5}^{2} \mbox{Tr}\Big({Y_u  Y_{u}^{\dagger}}\Big) \nonumber \\ 
  &-48 \lambda_6 \lambda_7 \mbox{Tr}\Big({Y_u  Y_{u}^{\dagger}}\Big) -24 \lambda_{7}^{2} \mbox{Tr}\Big({Y_u  Y_{u}^{\dagger}}\Big) -\frac{27}{2} \lambda_3 \mbox{Tr}\Big({Y_d  Y_{d}^{\dagger}  Y_d  Y_{d}^{\dagger}}\Big) -\frac{4}{5} g_{1}^{2} \mbox{Tr}\Big({Y_d  Y_{u}^{\dagger}  Y_u  Y_{d}^{\dagger}}\Big) \nonumber \\
  &-64 g_{3}^{2} \mbox{Tr}\Big({Y_d  Y_{u}^{\dagger}  Y_u  Y_{d}^{\dagger}}\Big) +15 \lambda_3 \mbox{Tr}\Big({Y_d  Y_{u}^{\dagger}  Y_u  Y_{d}^{\dagger}}\Big) -\frac{9}{2} \lambda_3 \mbox{Tr}\Big({Y_e  Y_{e}^{\dagger}  Y_e  Y_{e}^{\dagger}}\Big) \nonumber \\
  &-\frac{27}{2} \lambda_3 \mbox{Tr}\Big({Y_u  Y_{u}^{\dagger}  Y_u  Y_{u}^{\dagger}}\Big) +12 \mbox{Tr}\Big({Y_d  Y_{d}^{\dagger}  Y_d  Y_{u}^{\dagger}  Y_u  Y_{d}^{\dagger}}\Big) +24 \mbox{Tr}\Big({Y_d  Y_{u}^{\dagger}  Y_u  Y_{d}^{\dagger}  Y_d  Y_{d}^{\dagger}}\Big) \nonumber \\
  &+36 \mbox{Tr}\Big({Y_d  Y_{u}^{\dagger}  Y_u  Y_{u}^{\dagger}  Y_u  Y_{d}^{\dagger}}\Big) \\
  \beta_{\lambda_4}^{(1)} & =
 +\frac{9}{5} g_{1}^{2} g_{2}^{2} -\frac{9}{5} g_{1}^{2} \lambda_4 -9 g_{2}^{2} \lambda_4 +4 \lambda_1 \lambda_4 +4 \lambda_2 \lambda_4 +8 \lambda_3 \lambda_4 +4 \lambda_{4}^{2} +8 \lambda_{5}^{2} +10 \lambda_{6}^{2} +4 \lambda_6 \lambda_7 +10 \lambda_{7}^{2} \nonumber \\
  &+6 \lambda_4 \mbox{Tr}\Big({Y_d  Y_{d}^{\dagger}}\Big) +2 \lambda_4 \mbox{Tr}\Big({Y_e  Y_{e}^{\dagger}}\Big) +6 \lambda_4 \mbox{Tr}\Big({Y_u  Y_{u}^{\dagger}}\Big) +12 \mbox{Tr}\Big({Y_d  Y_{u}^{\dagger}  Y_u  Y_{d}^{\dagger}}\Big) \\
 \beta_{\lambda_4}^{(2)} & =
 -\frac{657}{50} g_{1}^{4} g_{2}^{2} -\frac{42}{5} g_{1}^{2} g_{2}^{4} +6 g_{1}^{2} g_{2}^{2} \lambda_1 +6 g_{1}^{2} g_{2}^{2} \lambda_2 +\frac{6}{5} g_{1}^{2} g_{2}^{2} \lambda_3 +\frac{1413}{200} g_{1}^{4} \lambda_4 +\frac{153}{20} g_{1}^{2} g_{2}^{2} \lambda_4 \nonumber \\
  &-\frac{231}{8} g_{2}^{4} \lambda_4 +\frac{24}{5} g_{1}^{2} \lambda_1 \lambda_4 -28 \lambda_{1}^{2} \lambda_4 +\frac{24}{5} g_{1}^{2} \lambda_2 \lambda_4 -28 \lambda_{2}^{2} \lambda_4 +\frac{12}{5} g_{1}^{2} \lambda_3 \lambda_4 +36 g_{2}^{2} \lambda_3 \lambda_4 \nonumber \\
  &-80 \lambda_1 \lambda_3 \lambda_4 -80 \lambda_2 \lambda_3 \lambda_4 -28 \lambda_{3}^{2} \lambda_4 +\frac{24}{5} g_{1}^{2} \lambda_{4}^{2} +18 g_{2}^{2} \lambda_{4}^{2} -40 \lambda_1 \lambda_{4}^{2} -40 \lambda_2 \lambda_{4}^{2} -28 \lambda_3 \lambda_{4}^{2} \nonumber \\
  &+\frac{48}{5} g_{1}^{2} \lambda_{5}^{2} +54 g_{2}^{2} \lambda_{5}^{2} -48 \lambda_1 \lambda_{5}^{2} -48 \lambda_2 \lambda_{5}^{2} -48 \lambda_3 \lambda_{5}^{2} -26 \lambda_4 \lambda_{5}^{2} +\frac{42}{5} g_{1}^{2} \lambda_{6}^{2} +54 g_{2}^{2} \lambda_{6}^{2} \nonumber \\
  &-148 \lambda_1 \lambda_{6}^{2} -20 \lambda_2 \lambda_{6}^{2} -72 \lambda_3 \lambda_{6}^{2} -68 \lambda_4 \lambda_{6}^{2} -80 \lambda_5 \lambda_{6}^{2} +\frac{24}{5} g_{1}^{2} \lambda_6 \lambda_7 -40 \lambda_1 \lambda_6 \lambda_7 -40 \lambda_2 \lambda_6 \lambda_7 \nonumber \\
  &-80 \lambda_3 \lambda_6 \lambda_7 -160 \lambda_4 \lambda_6 \lambda_7 -96 \lambda_5 \lambda_6 \lambda_7 +\frac{42}{5} g_{1}^{2} \lambda_{7}^{2} +54 g_{2}^{2} \lambda_{7}^{2} -20 \lambda_1 \lambda_{7}^{2} -148 \lambda_2 \lambda_{7}^{2} \nonumber \\
  &-72 \lambda_3 \lambda_{7}^{2} -68 \lambda_4 \lambda_{7}^{2} -80 \lambda_5 \lambda_{7}^{2} +\frac{1}{20} \Big(5 \Big(16 \Big(10 g_{3}^{2} \lambda_4 -3 \Big(2 \lambda_1 \lambda_4  + 2 \lambda_3 \lambda_4  + 2 \lambda_{5}^{2} \nonumber \\ 
  &+ 5 \lambda_{6}^{2}  + \lambda_6 \lambda_7  + \lambda_{4}^{2}\Big)\Big) + 45 g_{2}^{2} \lambda_4 \Big) + g_{1}^{2} \Big(108 g_{2}^{2}  + 25 \lambda_4 \Big)\Big)\mbox{Tr}\Big({Y_d  Y_{d}^{\dagger}}\Big) +\frac{1}{20} \Big(3 g_{1}^{2} \Big(25 \lambda_4 \nonumber \\  
  &+ 44 g_{2}^{2} \Big) + 5 \Big(15 g_{2}^{2} \lambda_4  -16 \Big(2 \lambda_1 \lambda_4  + 2 \lambda_3 \lambda_4  + 2 \lambda_{5}^{2}  + 5 \lambda_{6}^{2}  + \lambda_6 \lambda_7  + \lambda_{4}^{2}\Big)\Big)\Big)\mbox{Tr}\Big({Y_e  Y_{e}^{\dagger}}\Big) \nonumber \\
  &+\frac{63}{5} g_{1}^{2} g_{2}^{2} \mbox{Tr}\Big({Y_u  Y_{u}^{\dagger}}\Big) +\frac{17}{4} g_{1}^{2} \lambda_4 \mbox{Tr}\Big({Y_u  Y_{u}^{\dagger}}\Big) +\frac{45}{4} g_{2}^{2} \lambda_4 \mbox{Tr}\Big({Y_u  Y_{u}^{\dagger}}\Big) +40 g_{3}^{2} \lambda_4 \mbox{Tr}\Big({Y_u  Y_{u}^{\dagger}}\Big) \nonumber \\
  &-24 \lambda_2 \lambda_4 \mbox{Tr}\Big({Y_u  Y_{u}^{\dagger}}\Big) -24 \lambda_3 \lambda_4 \mbox{Tr}\Big({Y_u  Y_{u}^{\dagger}}\Big) -12 \lambda_{4}^{2} \mbox{Tr}\Big({Y_u  Y_{u}^{\dagger}}\Big) -24 \lambda_{5}^{2} \mbox{Tr}\Big({Y_u  Y_{u}^{\dagger}}\Big) \nonumber \\
  &-12 \lambda_6 \lambda_7 \mbox{Tr}\Big({Y_u  Y_{u}^{\dagger}}\Big) -60 \lambda_{7}^{2} \mbox{Tr}\Big({Y_u  Y_{u}^{\dagger}}\Big) -\frac{27}{2} \lambda_4 \mbox{Tr}\Big({Y_d  Y_{d}^{\dagger}  Y_d  Y_{d}^{\dagger}}\Big) +\frac{4}{5} g_{1}^{2} \mbox{Tr}\Big({Y_d  Y_{u}^{\dagger}  Y_u  Y_{d}^{\dagger}}\Big) \nonumber \\
  &+64 g_{3}^{2} \mbox{Tr}\Big({Y_d  Y_{u}^{\dagger}  Y_u  Y_{d}^{\dagger}}\Big) -24 \lambda_3 \mbox{Tr}\Big({Y_d  Y_{u}^{\dagger}  Y_u  Y_{d}^{\dagger}}\Big) -33 \lambda_4 \mbox{Tr}\Big({Y_d  Y_{u}^{\dagger}  Y_u  Y_{d}^{\dagger}}\Big) \nonumber \\
  &-\frac{9}{2} \lambda_4 \mbox{Tr}\Big({Y_e  Y_{e}^{\dagger}  Y_e  Y_{e}^{\dagger}}\Big) -\frac{27}{2} \lambda_4 \mbox{Tr}\Big({Y_u  Y_{u}^{\dagger}  Y_u  Y_{u}^{\dagger}}\Big) -12 \mbox{Tr}\Big({Y_d  Y_{d}^{\dagger}  Y_d  Y_{u}^{\dagger}  Y_u  Y_{d}^{\dagger}}\Big) \nonumber \\
  &-12 \mbox{Tr}\Big({Y_d  Y_{u}^{\dagger}  Y_u  Y_{d}^{\dagger}  Y_d  Y_{d}^{\dagger}}\Big)-24 \mbox{Tr}\Big({Y_d  Y_{u}^{\dagger}  Y_u  Y_{u}^{\dagger}  Y_u  Y_{d}^{\dagger}}\Big) \\
  \beta_{\lambda_5}^{(1)} & =
  -\frac{9}{5} g_{1}^{2} \lambda_5 -9 g_{2}^{2} \lambda_5 +4 \lambda_1 \lambda_5 +4 \lambda_2 \lambda_5 +8 \lambda_3 \lambda_5 +12 \lambda_4 \lambda_5 +10 \lambda_{6}^{2} +4 \lambda_6 \lambda_7 +10 \lambda_{7}^{2} \nonumber \\ 
  &+6 \lambda_5 \mbox{Tr}\Big({Y_d  Y_{d}^{\dagger}}\Big) +2 \lambda_5 \mbox{Tr}\Big({Y_e  Y_{e}^{\dagger}}\Big) +6 \lambda_5 \mbox{Tr}\Big({Y_u  Y_{u}^{\dagger}}\Big) \\
  \beta_{\lambda_5}^{(2)} & =
  +\frac{1413}{200} g_{1}^{4} \lambda_5 +\frac{57}{20} g_{1}^{2} g_{2}^{2} \lambda_5 -\frac{231}{8} g_{2}^{4} \lambda_5 -\frac{12}{5} g_{1}^{2} \lambda_1 \lambda_5 -28 \lambda_{1}^{2} \lambda_5 -\frac{12}{5} g_{1}^{2} \lambda_2 \lambda_5 -28 \lambda_{2}^{2} \lambda_5 \nonumber \\
   &+\frac{48}{5} g_{1}^{2} \lambda_3 \lambda_5 +36 g_{2}^{2} \lambda_3 \lambda_5 -80 \lambda_1 \lambda_3 \lambda_5 -80 \lambda_2 \lambda_3 \lambda_5 -28 \lambda_{3}^{2} \lambda_5 +\frac{72}{5} g_{1}^{2} \lambda_4 \lambda_5 +72 g_{2}^{2} \lambda_4 \lambda_5 \nonumber \\
   &-88 \lambda_1 \lambda_4 \lambda_5 -88 \lambda_2 \lambda_4 \lambda_5 -76 \lambda_3 \lambda_4 \lambda_5 -32 \lambda_{4}^{2} \lambda_5 +6 \lambda_{5}^{3} +12 g_{1}^{2} \lambda_{6}^{2} +54 g_{2}^{2} \lambda_{6}^{2} -148 \lambda_1 \lambda_{6}^{2} \nonumber \\
   &-20 \lambda_2 \lambda_{6}^{2} -72 \lambda_3 \lambda_{6}^{2} -76 \lambda_4 \lambda_{6}^{2} -72 \lambda_5 \lambda_{6}^{2} -\frac{12}{5} g_{1}^{2} \lambda_6 \lambda_7 -40 \lambda_1 \lambda_6 \lambda_7 -40 \lambda_2 \lambda_6 \lambda_7 \nonumber \\
   &-80 \lambda_3 \lambda_6 \lambda_7 -88 \lambda_4 \lambda_6 \lambda_7 -168 \lambda_5 \lambda_6 \lambda_7 +12 g_{1}^{2} \lambda_{7}^{2} +54 g_{2}^{2} \lambda_{7}^{2} -20 \lambda_1 \lambda_{7}^{2} -148 \lambda_2 \lambda_{7}^{2} \nonumber \\
   &-72 \lambda_3 \lambda_{7}^{2} -76 \lambda_4 \lambda_{7}^{2} -72 \lambda_5 \lambda_{7}^{2} \nonumber \\
   &+\frac{1}{4} \Big(16 \Big(10 g_{3}^{2} \lambda_5  -3 \Big(2 \lambda_1 \lambda_5  + 2 \lambda_3 \lambda_5  + 3 \lambda_4 \lambda_5  + 5 \lambda_{6}^{2}  + \lambda_6 \lambda_7 \Big)\Big) + 45 g_{2}^{2} \lambda_5  + 5 g_{1}^{2} \lambda_5 \Big)\mbox{Tr}\Big({Y_d  Y_{d}^{\dagger}}\Big) \nonumber \\
   &+\frac{1}{4} \Big(15 g_{1}^{2} \lambda_5  + 15 g_{2}^{2} \lambda_5  -16 \Big(2 \lambda_1 \lambda_5  + 2 \lambda_3 \lambda_5  + 3 \lambda_4 \lambda_5  + 5 \lambda_{6}^{2}  + \lambda_6 \lambda_7 \Big)\Big)\mbox{Tr}\Big({Y_e  Y_{e}^{\dagger}}\Big) \nonumber \\
   &+\frac{17}{4} g_{1}^{2} \lambda_5 \mbox{Tr}\Big({Y_u  Y_{u}^{\dagger}}\Big) +\frac{45}{4} g_{2}^{2} \lambda_5 \mbox{Tr}\Big({Y_u  Y_{u}^{\dagger}}\Big) +40 g_{3}^{2} \lambda_5 \mbox{Tr}\Big({Y_u  Y_{u}^{\dagger}}\Big) -24 \lambda_2 \lambda_5 \mbox{Tr}\Big({Y_u  Y_{u}^{\dagger}}\Big) \nonumber \\
   &-24 \lambda_3 \lambda_5 \mbox{Tr}\Big({Y_u  Y_{u}^{\dagger}}\Big) -36 \lambda_4 \lambda_5 \mbox{Tr}\Big({Y_u  Y_{u}^{\dagger}}\Big) -12 \lambda_6 \lambda_7 \mbox{Tr}\Big({Y_u  Y_{u}^{\dagger}}\Big) -60 \lambda_{7}^{2} \mbox{Tr}\Big({Y_u  Y_{u}^{\dagger}}\Big) \nonumber \\
   &-\frac{3}{2} \lambda_5 \mbox{Tr}\Big({Y_d  Y_{d}^{\dagger}  Y_d  Y_{d}^{\dagger}}\Big) -33 \lambda_5 \mbox{Tr}\Big({Y_d  Y_{u}^{\dagger}  Y_u  Y_{d}^{\dagger}}\Big) -\frac{1}{2} \lambda_5 \mbox{Tr}\Big({Y_e  Y_{e}^{\dagger}  Y_e  Y_{e}^{\dagger}}\Big) -\frac{3}{2} \lambda_5 \mbox{Tr}\Big({Y_u  Y_{u}^{\dagger}  Y_u  Y_{u}^{\dagger}}\Big) \\
\beta_{\lambda_6}^{(1)} & =
-\frac{9}{5} g_{1}^{2} \lambda_6 -9 g_{2}^{2} \lambda_6 +24 \lambda_1 \lambda_6 +6 \lambda_3 \lambda_6 +8 \lambda_4 \lambda_6 +10 \lambda_5 \lambda_6 +6 \lambda_3 \lambda_7 +4 \lambda_4 \lambda_7 +2 \lambda_5 \lambda_7\nonumber \\ 
&+9 \lambda_6 \mbox{Tr}\Big({Y_d  Y_{d}^{\dagger}}\Big) +3 \lambda_6 \mbox{Tr}\Big({Y_e  Y_{e}^{\dagger}}\Big) +3 \lambda_6 \mbox{Tr}\Big({Y_u  Y_{u}^{\dagger}}\Big) \\
\beta_{\lambda_6}^{(2)} & =
+\frac{1683}{200} g_{1}^{4} \lambda_6 +\frac{87}{20} g_{1}^{2} g_{2}^{2} \lambda_6 -\frac{141}{8} g_{2}^{4} \lambda_6 +\frac{108}{5} g_{1}^{2} \lambda_1 \lambda_6 +108 g_{2}^{2} \lambda_1 \lambda_6 -318 \lambda_{1}^{2} \lambda_6 +6 \lambda_{2}^{2} \lambda_6 \nonumber \\
 &+\frac{18}{5} g_{1}^{2} \lambda_3 \lambda_6 +18 g_{2}^{2} \lambda_3 \lambda_6 -132 \lambda_1 \lambda_3 \lambda_6 -36 \lambda_2 \lambda_3 \lambda_6 -32 \lambda_{3}^{2} \lambda_6 +6 g_{1}^{2} \lambda_4 \lambda_6 +36 g_{2}^{2} \lambda_4 \lambda_6 \nonumber \\
 &-140 \lambda_1 \lambda_4 \lambda_6 -28 \lambda_2 \lambda_4 \lambda_6 -68 \lambda_3 \lambda_4 \lambda_6 -34 \lambda_{4}^{2} \lambda_6 +12 g_{1}^{2} \lambda_5 \lambda_6 +54 g_{2}^{2} \lambda_5 \lambda_6 -148 \lambda_1 \lambda_5 \lambda_6 \nonumber \\
 &-20 \lambda_2 \lambda_5 \lambda_6 -72 \lambda_3 \lambda_5 \lambda_6 -76 \lambda_4 \lambda_5 \lambda_6 -36 \lambda_{5}^{2} \lambda_6 -111 \lambda_{6}^{3} +\frac{27}{20} g_{1}^{4} \lambda_7 +\frac{3}{2} g_{1}^{2} g_{2}^{2} \lambda_7 +\frac{45}{4} g_{2}^{4} \lambda_7 \nonumber \\
 &+\frac{36}{5} g_{1}^{2} \lambda_3 \lambda_7 +36 g_{2}^{2} \lambda_3 \lambda_7 -36 \lambda_1 \lambda_3 \lambda_7 -36 \lambda_2 \lambda_3 \lambda_7 -36 \lambda_{3}^{2} \lambda_7 +\frac{24}{5} g_{1}^{2} \lambda_4 \lambda_7 +18 g_{2}^{2} \lambda_4 \lambda_7 \nonumber \\
 &-28 \lambda_1 \lambda_4 \lambda_7 -28 \lambda_2 \lambda_4 \lambda_7 -56 \lambda_3 \lambda_4 \lambda_7 -34 \lambda_{4}^{2} \lambda_7 -\frac{6}{5} g_{1}^{2} \lambda_5 \lambda_7 -20 \lambda_1 \lambda_5 \lambda_7 -20 \lambda_2 \lambda_5 \lambda_7 \nonumber \\
 &-40 \lambda_3 \lambda_5 \lambda_7 -44 \lambda_4 \lambda_5 \lambda_7 -42 \lambda_{5}^{2} \lambda_7 -126 \lambda_{6}^{2} \lambda_7 -33 \lambda_6 \lambda_{7}^{2} -42 \lambda_{7}^{3} \nonumber \\
 &+\frac{3}{8} \Big(16 \Big(10 g_{3}^{2}  -24 \lambda_1  -3 \lambda_3  -4 \lambda_4  -5 \lambda_5 \Big) + 45 g_{2}^{2}  + 5 g_{1}^{2} \Big)\lambda_6 \mbox{Tr}\Big({Y_d  Y_{d}^{\dagger}}\Big) \nonumber \\
 &+\frac{1}{8} \Big(-16 \Big(24 \lambda_1  + 3 \lambda_3  + 4 \lambda_4  + 5 \lambda_5 \Big) + 45 g_{1}^{2}  + 45 g_{2}^{2} \Big)\lambda_6 \mbox{Tr}\Big({Y_e  Y_{e}^{\dagger}}\Big) +\frac{17}{8} g_{1}^{2} \lambda_6 \mbox{Tr}\Big({Y_u  Y_{u}^{\dagger}}\Big) \nonumber \\
 &+\frac{45}{8} g_{2}^{2} \lambda_6 \mbox{Tr}\Big({Y_u  Y_{u}^{\dagger}}\Big) +20 g_{3}^{2} \lambda_6 \mbox{Tr}\Big({Y_u  Y_{u}^{\dagger}}\Big) -18 \lambda_3 \lambda_6 \mbox{Tr}\Big({Y_u  Y_{u}^{\dagger}}\Big) -24 \lambda_4 \lambda_6 \mbox{Tr}\Big({Y_u  Y_{u}^{\dagger}}\Big) \nonumber \\
 &-30 \lambda_5 \lambda_6 \mbox{Tr}\Big({Y_u  Y_{u}^{\dagger}}\Big) -36 \lambda_3 \lambda_7 \mbox{Tr}\Big({Y_u  Y_{u}^{\dagger}}\Big) -24 \lambda_4 \lambda_7 \mbox{Tr}\Big({Y_u  Y_{u}^{\dagger}}\Big) -12 \lambda_5 \lambda_7 \mbox{Tr}\Big({Y_u  Y_{u}^{\dagger}}\Big) \nonumber \\
 &-\frac{33}{4} \lambda_6 \mbox{Tr}\Big({Y_d  Y_{d}^{\dagger}  Y_d  Y_{d}^{\dagger}}\Big) -21 \lambda_6 \mbox{Tr}\Big({Y_d  Y_{u}^{\dagger}  Y_u  Y_{d}^{\dagger}}\Big) -\frac{11}{4} \lambda_6 \mbox{Tr}\Big({Y_e  Y_{e}^{\dagger}  Y_e  Y_{e}^{\dagger}}\Big) \nonumber \\ 
 &-\frac{27}{4} \lambda_6 \mbox{Tr}\Big({Y_u  Y_{u}^{\dagger}  Y_u  Y_{u}^{\dagger}}\Big) \\
 \beta_{\lambda_7}^{(1)} & =
+6 \lambda_3 \lambda_6 +4 \lambda_4 \lambda_6 +2 \lambda_5 \lambda_6 -\frac{9}{5} g_{1}^{2} \lambda_7 -9 g_{2}^{2} \lambda_7 +24 \lambda_2 \lambda_7 +6 \lambda_3 \lambda_7 +8 \lambda_4 \lambda_7 +10 \lambda_5 \lambda_7 \nonumber \\ 
&+3 \lambda_7 \mbox{Tr}\Big({Y_d  Y_{d}^{\dagger}}\Big) +\lambda_7 \mbox{Tr}\Big({Y_e  Y_{e}^{\dagger}}\Big) +9 \lambda_7 \mbox{Tr}\Big({Y_u  Y_{u}^{\dagger}}\Big) \\
\beta_{\lambda_7}^{(2)} & =
+\frac{27}{20} g_{1}^{4} \lambda_6 +\frac{3}{2} g_{1}^{2} g_{2}^{2} \lambda_6 +\frac{45}{4} g_{2}^{4} \lambda_6 +\frac{36}{5} g_{1}^{2} \lambda_3 \lambda_6 +36 g_{2}^{2} \lambda_3 \lambda_6 -36 \lambda_1 \lambda_3 \lambda_6 -36 \lambda_2 \lambda_3 \lambda_6 \nonumber \\
 &-36 \lambda_{3}^{2} \lambda_6 +\frac{24}{5} g_{1}^{2} \lambda_4 \lambda_6 +18 g_{2}^{2} \lambda_4 \lambda_6 -28 \lambda_1 \lambda_4 \lambda_6 -28 \lambda_2 \lambda_4 \lambda_6 -56 \lambda_3 \lambda_4 \lambda_6 -34 \lambda_{4}^{2} \lambda_6 \nonumber \\
 &-\frac{6}{5} g_{1}^{2} \lambda_5 \lambda_6 -20 \lambda_1 \lambda_5 \lambda_6 -20 \lambda_2 \lambda_5 \lambda_6 -40 \lambda_3 \lambda_5 \lambda_6 -44 \lambda_4 \lambda_5 \lambda_6 -42 \lambda_{5}^{2} \lambda_6 -42 \lambda_{6}^{3} +\frac{1683}{200} g_{1}^{4} \lambda_7 \nonumber \\
 &+\frac{87}{20} g_{1}^{2} g_{2}^{2} \lambda_7 -\frac{141}{8} g_{2}^{4} \lambda_7 +6 \lambda_{1}^{2} \lambda_7 +\frac{108}{5} g_{1}^{2} \lambda_2 \lambda_7 +108 g_{2}^{2} \lambda_2 \lambda_7 -318 \lambda_{2}^{2} \lambda_7 +\frac{18}{5} g_{1}^{2} \lambda_3 \lambda_7 \nonumber \\
 &+18 g_{2}^{2} \lambda_3 \lambda_7 -36 \lambda_1 \lambda_3 \lambda_7 -132 \lambda_2 \lambda_3 \lambda_7 -32 \lambda_{3}^{2} \lambda_7 +6 g_{1}^{2} \lambda_4 \lambda_7 +36 g_{2}^{2} \lambda_4 \lambda_7 -28 \lambda_1 \lambda_4 \lambda_7 \nonumber \\
 &-140 \lambda_2 \lambda_4 \lambda_7 -68 \lambda_3 \lambda_4 \lambda_7 -34 \lambda_{4}^{2} \lambda_7 +12 g_{1}^{2} \lambda_5 \lambda_7 +54 g_{2}^{2} \lambda_5 \lambda_7 -20 \lambda_1 \lambda_5 \lambda_7 -148 \lambda_2 \lambda_5 \lambda_7 \nonumber \\
 &-72 \lambda_3 \lambda_5 \lambda_7 -76 \lambda_4 \lambda_5 \lambda_7 -36 \lambda_{5}^{2} \lambda_7 -33 \lambda_{6}^{2} \lambda_7 -126 \lambda_6 \lambda_{7}^{2} -111 \lambda_{7}^{3} -\frac{1}{8} \Big(144 \lambda_3 \Big(2 \lambda_6  + \lambda_7\Big) \nonumber \\ 
 &-160 g_{3}^{2} \lambda_7  + 192 \lambda_4 \Big(\lambda_6 + \lambda_7\Big) + 240 \lambda_5 \lambda_7  -45 g_{2}^{2} \lambda_7  -5 g_{1}^{2} \lambda_7  + 96 \lambda_5 \lambda_6 \Big)\mbox{Tr}\Big({Y_d  Y_{d}^{\dagger}}\Big) \nonumber \\
 &-\frac{1}{8} \Big(-15 g_{1}^{2} \lambda_7  -15 g_{2}^{2} \lambda_7  + 32 \lambda_5 \lambda_6  + 48 \lambda_3 \Big(2 \lambda_6  + \lambda_7\Big) + 64 \lambda_4 \Big(\lambda_6 + \lambda_7\Big) + 80 \lambda_5 \lambda_7 \Big)\mbox{Tr}\Big({Y_e  Y_{e}^{\dagger}}\Big) \nonumber \\
 &+\frac{51}{8} g_{1}^{2} \lambda_7 \mbox{Tr}\Big({Y_u  Y_{u}^{\dagger}}\Big) +\frac{135}{8} g_{2}^{2} \lambda_7 \mbox{Tr}\Big({Y_u  Y_{u}^{\dagger}}\Big) +60 g_{3}^{2} \lambda_7 \mbox{Tr}\Big({Y_u  Y_{u}^{\dagger}}\Big) -144 \lambda_2 \lambda_7 \mbox{Tr}\Big({Y_u  Y_{u}^{\dagger}}\Big) \nonumber \\
 &-18 \lambda_3 \lambda_7 \mbox{Tr}\Big({Y_u  Y_{u}^{\dagger}}\Big) -24 \lambda_4 \lambda_7 \mbox{Tr}\Big({Y_u  Y_{u}^{\dagger}}\Big) -30 \lambda_5 \lambda_7 \mbox{Tr}\Big({Y_u  Y_{u}^{\dagger}}\Big) -\frac{27}{4} \lambda_7 \mbox{Tr}\Big({Y_d  Y_{d}^{\dagger}  Y_d  Y_{d}^{\dagger}}\Big) \nonumber \\
 &-21 \lambda_7 \mbox{Tr}\Big({Y_d  Y_{u}^{\dagger}  Y_u  Y_{d}^{\dagger}}\Big) -\frac{9}{4} \lambda_7 \mbox{Tr}\Big({Y_e  Y_{e}^{\dagger}  Y_e  Y_{e}^{\dagger}}\Big) -\frac{33}{4} \lambda_7 \mbox{Tr}\Big({Y_u  Y_{u}^{\dagger}  Y_u  Y_{u}^{\dagger}}\Big)
\end{align}}